\documentclass{article}

\usepackage{PRIMEarxiv}
\usepackage[utf8]{inputenc}
\usepackage[T1]{fontenc}
\usepackage{hyperref}
\usepackage{url}
\usepackage{booktabs}
\usepackage{amsfonts}
\usepackage{amsmath}
\usepackage{microtype}
\usepackage{graphicx}
\usepackage{array}
\usepackage{float}
\usepackage{longtable}
\graphicspath{{media/}}

\hypersetup{colorlinks=true,linkcolor=blue,citecolor=blue,urlcolor=blue}

\title{Multi-view fnirs identifies neurophysiological subtypes in post-stroke aphasia}

\author{\parbox{0.96\textwidth}{\centering
Zijian Kang$^{1,\dagger}$, Bingbing Lin$^{2,3,\dagger}$, Jinglei Ni$^{2,3}$, Meifang Tian$^{2,3}$, Wai Ting Siok$^1$, Jia Huang$^{2,3}$, and Nizhuan Wang$^1$\\[0.6em]
\small $^1$Department of Language Science and Technology, The Hong Kong Polytechnic University, Hung Hom, Kowloon, Hong Kong Special Administrative Region, China.\\
\small $^2$College of Rehabilitation Medicine, Fujian University of Traditional Chinese Medicine, Fuzhou, China.\\
\small $^3$National-Local Joint Engineering Research Center of Rehabilitation Medicine Technology, Fujian University of Traditional Chinese Medicine, Fuzhou, China.\\[0.4em]
\small $^\dagger$These authors contributed equally to this work.\\[0.6em]
\small Correspondence: Nizhuan Wang, \href{mailto:wangnizhuan1120@gmail.com}{wangnizhuan1120@gmail.com}; Jia Huang, \href{mailto:jasmine1874@163.com}{jasmine1874@163.com}
}}

\begin{document}
\maketitle

\begin{abstract}
Post-stroke aphasia is clinically heterogeneous, and conventional syndrome-based classifications may not capture underlying differences in brain functional organisation. We investigated whether integrated resting-state and task-based functional near-infrared spectroscopy (fNIRS) combining with artificial intelligence (AI) methods could identify neurophysiological subgroups and provide characterisation complementary to conventional clinical and behavioural assessment.

Seventy-nine patients with post-stroke aphasia (mean age, 61 \(\pm\) 11 years; range, 36--83; 65 male and 14 female patients) underwent unsupervised clustering based on three feature views derived from fNIRS, i.e., resting-state intrinsic activity and connectivity, activation maps during picture naming and phonemic and semantic fluency, and connectivity reconfiguration between resting and task conditions. Similarity network fusion and spectral clustering were used to integrate these views and identify subgroups. Seventeen healthy controls (mean age, 61 \(\pm\) 8 years; 7 male and 10 female controls) served as a reference group for neurophysiological profiling, while clinical and behavioural measures were not used for clustering.

Three clusters of 32, 17 and 30 patients were identified; this solution had the largest eigengap and was the only candidate significant after permutation testing and Holm correction (adjusted \emph{P} = 0.001). It was stable under patient subsampling (mean adjusted Rand index, 0.773) but depended most on the resting-state intrinsic view, whose removal lowered agreement to 0.217. Clusters differed in 19 of 22 regional measures of intrinsic activity and all six resting-state connectivity summaries, but in only one of 66 task-activation features and none of 12 reconfiguration summaries. The first cluster showed the highest right anterior prefrontal intrinsic activity, the second distributed frontal, sensorimotor and parietal alterations, and the third the greatest reduction in mean interhemispheric connectivity relative to healthy controls.
Although age differed across clusters, the between-cluster connectivity difference persisted after age adjustment. Cluster membership was not significantly associated with conventional aphasia subtype (Cramér's \emph{V} = 0.287), and baseline clinical language scores and behavioural task performance did not differ significantly across clusters.

These findings identify three candidate neurophysiological subgroups in post-stroke aphasia, distinguished mainly by intrinsic brain activity and resting-state functional connectivity, that did not clearly align with conventional aphasia classifications or baseline clinical measures in this cohort. Multi-view functional near-infrared spectroscopy analysis may provide a complementary framework for characterizing patient heterogeneity where the independent replication and longitudinal follow-up are needed to establish its reproducibility and clinical utility.
\end{abstract}


\section{Introduction}\label{introduction}

Post-stroke aphasia is an acquired language disorder caused by cerebrovascular injury, affecting language production, comprehension or both. Its presentation varies across speech fluency, auditory comprehension, repetition, naming, reading and writing \cite{ref01}. Aphasia is a common and consequential complication of stroke: a systematic review of 50 studies reported median frequencies of 30\% in acute-care and 34\% in rehabilitation settings, and found that patients with aphasia had greater disability and returned home less frequently than those without aphasia \cite{ref02}. It can also restrict everyday participation and reduce quality of life \cite{ref03}. Clinically, aphasia has traditionally been classified into syndromes such as anomic, Broca's, conduction, global, mixed transcortical, transcortical motor, transcortical sensory and Wernicke's aphasia, reflecting classical clinico-anatomical models of language \cite{ref04}. These syndromic distinctions are operationalised in standardised clinical assessments, including the Boston Diagnostic Aphasia Examination (BDAE) and the Western Aphasia Battery (WAB), which assess fluency, comprehension, repetition and other language domains \cite{ref05}.

Although conventional syndrome-based classifications remain useful for clinical description, their categorical structure may obscure substantial heterogeneity among individuals assigned to the same subtype and overlapping behavioural profiles across different subtypes \cite{ref06,ref07,ref08}. Previous studies have reported limited agreement between subtype assignments derived from the WAB and BDAE, poor correspondence between conventional subtypes and data-derived groupings, and aphasic profiles that cannot be assigned to a conventional category \cite{ref09,ref10,ref11}. Dimensional approaches have identified graded behavioral dimensions involving phonology, semantics, fluency and executive abilities \cite{ref12,ref13}. while community detection applied to patients' behavioural profiles has revealed clusters that do not map directly onto conventional aphasia subtypes \cite{ref14}. Lesion mapping has related these dimensions and clusters to distinct patterns of structural damage, but they were derived primarily from behavioural measures rather than from functional brain activity or connectivity.

Functional neuroimaging studies have revealed substantial inter-individual variation in cortical activation and functional connectivity among patients with post-stroke aphasia. Meta-analytic evidence most consistently implicates spared left-hemisphere language regions, with limited support for differential recruitment of right-hemisphere or domain-general regions \cite{ref15}. At the regional level, resting-state haemodynamic timing shifts in intact tissue have been related to distinct language components, and task-evoked activation of the canonical semantic network accounts for variation in residual language function \cite{ref16,ref17}. At the network level, electrophysiological and haemodynamic connectivity have been associated with preserved language functions, shown longitudinally to reorganise during recovery according to lesion location, and found to vary with aphasia type and severity \cite{ref18,ref19,ref20}. Such neurophysiological measures may capture aspects of functional brain organisation not fully reflected in clinical language assessments. However, they have mainly been related to behavioral scores or predefined clinical or lesion-based groups, and rarely used to define patient subgroups.

Functional near-infrared spectroscopy (fNIRS) is a non-invasive optical neuroimaging technique that estimates cortical haemodynamic changes through variations in oxygenated and deoxygenated haemoglobin. Its portability and flexibility for monitoring cortical responses during overt language tasks make it suitable for clinical populations, although its measurements are largely restricted to superficial cortex and susceptible to motion and systemic physiological artifacts \cite{ref21,ref22}. In post-stroke aphasia, fNIRS has been used to examine task-evoked responses during semantic processing and picture naming, as well as treatment-related changes in speech-associated networks \cite{ref23,ref24,ref54}. Resting-state fNIRS has also been used to characterize intrinsic functional connectivity in acute and subacute aphasia and network properties in non-fluent aphasia \cite{ref25,ref26}. A recent intervention study acquired both resting-state and task-based measurements in the same cohort, illustrating the feasibility of combining the two conditions \cite{ref27}. These conditions capture complementary aspects of brain function: resting-state measurements characterize intrinsic functional organization without requiring task performance, whereas task-based measurements capture responses to defined linguistic demands. Contrasting functional connectivity between these conditions, and across different language-task contexts, may further capture state-dependent network reconfiguration. Integrating intrinsic activity and connectivity, task-evoked activation and connectivity reconfiguration may therefore provide a more comprehensive representation of inter-individual neurophysiological variation. However, previous fNIRS studies in aphasia have mainly focused on group-level differences or treatment effects, and whether such integrated profiles can define neurophysiological subgroups in a data-driven manner has not been systematically examined.

In this paper, we aimed to identify and characterize candidate neurophysiological subgroups of post-stroke aphasia using a multi-view fNIRS approach. The corresponding fNIRS data were acquired at rest and during three language tasks: picture naming, phonemic fluency and semantic fluency. From these recordings, we derived three feature views, namely resting-state intrinsic organization, multi-task activation and rest--task connectivity reconfiguration. These views were integrated using similarity network fusion, and patient clusters were identified using spectral clustering. Clinical and behavioral measures were not used for clustering, allowing their correspondence with the resulting clusters to be examined independently. Here, we addressed three questions: (i) whether multi-view fNIRS could identify stable subtype patient clusters; (ii) how the neurophysiological profiles of these clusters differed from those of healthy controls and from one another; and (iii) whether cluster membership corresponded to conventional clinical classification, baseline language scores or behavioral task performance.

\section{Materials and methods}\label{materials-and-methods}

\subsection{Participants}\label{participants}

\subsubsection{Participant recruitment}\label{participant-recruitment}

Two groups of participants were enrolled: patients with post-stroke aphasia and healthy controls. Patients were recruited from inpatients at the Rehabilitation Hospital Affiliated with Fujian University of Traditional Chinese Medicine (Fuzhou, China) between May and December 2022. During the same period, healthy control volunteers were recruited from the community, with no history of cognitive impairment or language-related neurological disorders. All participants were native Mandarin Chinese speakers and completed fNIRS recordings under both resting-state and task-based conditions.

\subsubsection{Inclusion and exclusion criteria}\label{inclusion-and-exclusion-criteria}

The inclusion criteria were as follows: (i) a diagnosis of ischaemic, haemorrhagic or mixed stroke according to the International Classification of Diseases (ICD), 10th Revision (ICD-10; codes I63.902 for ischaemic and I61.902 for haemorrhagic stroke), with lesions located in the left cerebral hemisphere; (ii) a first-ever diagnosis of aphasia following stroke; (iii) right-handedness; (iv) age between 35 and 85 years; and (v) stroke onset within nine months of enrolment.

The exclusion criteria were as follows: (i) aphasia caused by conditions other than stroke; (ii) speech impairment attributable to dysarthria; (iii) severe auditory or visual impairment, or severe cognitive impairment, precluding participation in language assessment or fNIRS testing; (iv) severe post-stroke complications, including pulmonary infection, shoulder--hand syndrome or lower-limb venous thromboembolism; (v) severe cardiac disease or failure, hepatic or renal failure, malignant tumours or gastrointestinal bleeding; and (vi) a history of epilepsy or depressive disorder.

\begin{table}[t]
\centering
\caption{Demographic and clinical characteristics by traditional WAB subtype}
\label{tab:1}
\scriptsize
\setlength{\tabcolsep}{1.8pt}
\renewcommand{\arraystretch}{1.15}
\begin{tabular}{>{\raggedright\arraybackslash}p{2.75cm}*{7}{>{\centering\arraybackslash}p{1.55cm}}}
\toprule
\textbf{WAB diagnosis} & \textbf{n} & \textbf{\shortstack{Age,\\years}} & \textbf{\shortstack{Female/\\male, n}} & \textbf{\shortstack{Education,\\years}} & \textbf{\shortstack{Days post\\onset}} & \textbf{\shortstack{WAB\\baseline\\score}} & \textbf{\shortstack{BDAE\\baseline\\score}} \\
\midrule
Anomic & 44 & \shortstack{61\\(43--79)} & 6/38 & \shortstack{9.68\\(0--16)} & \shortstack{40.07\\(0--171)} & \shortstack{364.02\\(290--415)} & \shortstack{3.30\\(2--5)} \\
Broca's & 7 & \shortstack{50\\(36--71)} & 1/6 & \shortstack{10.43\\(6--16)} & \shortstack{93.57\\(16--205)} & \shortstack{161.93\\(24.5--317)} & \shortstack{1.43\\(1--2)} \\
Conduction & 1 & 62 & 0/1 & 9 & 16 & 285 & 3 \\
Global & 9 & \shortstack{69\\(57--76)} & 4/5 & \shortstack{5.78\\(0--16)} & \shortstack{72.89\\(13--221)} & \shortstack{28.94\\(0--109)} & \shortstack{1.11\\(1--2)} \\
Mixed transcortical & 2 & \shortstack{65\\(62--67)} & 0/2 & \shortstack{9.00\\(9--9)} & \shortstack{64.00\\(13--115)} & \shortstack{154.00\\(143--165)} & \shortstack{2\\(2--2)} \\
Transcortical motor & 4 & \shortstack{64\\(56--71)} & 0/4 & \shortstack{7.50\\(6--9)} & \shortstack{48.75\\(20--87)} & \shortstack{327.00\\(316--350)} & \shortstack{2.50\\(1--3)} \\
Transcortical sensory & 3 & \shortstack{65\\(55--83)} & 0/3 & \shortstack{10.00\\(6--12)} & \shortstack{13.67\\(6--19)} & \shortstack{204.33\\(131--286)} & \shortstack{1.67\\(1--2)} \\
Wernicke's & 9 & \shortstack{60\\(37--79)} & 3/6 & \shortstack{11.11\\(9--16)} & \shortstack{43.89\\(7--140)} & \shortstack{177.28\\(107--259)} & \shortstack{1.78\\(1--3)} \\
\textbf{Overall} & \textbf{79} & \textbf{\shortstack{61\\(36--83)}} & \textbf{14/65} & \textbf{\shortstack{9.34\\(0--16)}} & \textbf{\shortstack{48.72\\(0--221)}} & \textbf{\shortstack{272.41\\(0--415)}} & \textbf{\shortstack{2.57\\(1--5)}} \\
\bottomrule
\end{tabular}
\vspace{0.6em}
\begin{minipage}{\textwidth}
\footnotesize
Values are presented as mean (range), except sex, which is presented as female/male counts.

The table includes 79 patients with matched traditional WAB subtype and clinical data. Disease duration is reported in days as recorded in the source data.

WAB baseline is the supplied 0-415 score and not WAB Aphasia Quotient. BDAE baseline is the recorded 1-5 score.

BDAE, Boston Diagnostic Aphasia Examination; WAB, Western Aphasia Battery.
\end{minipage}
\end{table}

\subsubsection{Demographic and clinical characteristics}\label{demographic-and-clinical-characteristics}

The study included 79 patients with post-stroke aphasia (male/female: 65/14; mean age \(\pm\) standard deviation (SD): 61 \(\pm\) 11 years) and 17 healthy controls (male/female: 7/10; mean age \(\pm\) SD: 61 \(\pm\) 8 years). Healthy-control demographics were available only as group-level data. Patients represented eight traditional WAB-defined aphasia subtypes: anomic, Broca's, conduction, global, mixed transcortical, transcortical motor, transcortical sensory, and Wernicke's aphasia \cite{ref28}. Patient demographic and clinical characteristics by aphasia subtype are summarized in Table~\ref{tab:1}.

\subsubsection{Ethical approval}\label{ethical-approval}

The study was approved by the Ethics Committee of the Rehabilitation Hospital Affiliated with Fujian University of Traditional Chinese Medicine (approval No.~2021KY-001-01). All study procedures were conducted in accordance with the Declaration of Helsinki. Written informed consent was obtained from all participants or their authorised representatives prior to enrolment.

\subsection{Experiments}\label{experiments}

\subsubsection{fNIRS data acquisition}\label{fnirs-data-acquisition}

fNIRS data were acquired using a LABNIRS system (Shimadzu Corporation, Japan), equipped with 32 light emitters and 33 detectors, forming 110 measurement channels covering cortical regions associated with language function. The inter-optode distance was 3 cm, and the sampling frequency was 10.101 Hz. The system was used to monitor changes in oxygenated haemoglobin (HbO) concentration. The three-dimensional positions of the optodes were recorded using a three-dimensional digitiser, from which the corresponding channel coordinates were derived and transformed into Montreal Neurological Institute (MNI) space \cite{ref29}. The MNI coordinates of all 110 measurement channels are provided in Appendix Table~\ref{tab:supp-mni-coordinates}.

\subsubsection{Experimental paradigms}\label{experimental-paradigms}

\paragraph{Resting-state paradigm}\label{resting-state-paradigm}

Participants were instructed to sit quietly with their eyes closed during a continuous 3-min resting-state recording.

\paragraph{Picture-naming task}\label{picture-naming-task}

\begin{figure}[t]
\centering
\includegraphics[width=\textwidth,height=0.65\textheight,keepaspectratio]{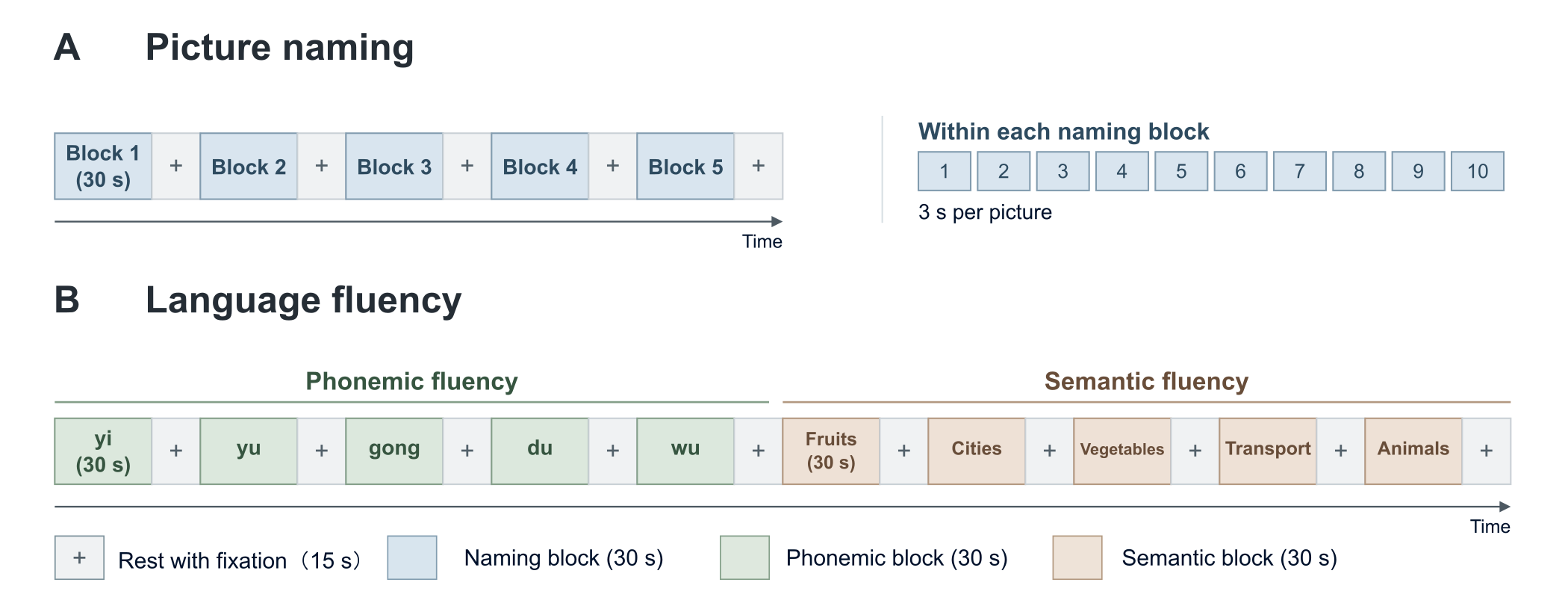}
\caption{\textbf{Picture naming and language fluency paradigms.} \textbf{(A)} In five 30-s blocks, participants named 10 pictures aloud, each presented for 3 s. \textbf{(B)} Five 30-s phonemic fluency blocks preceded five 30-s semantic fluency blocks in a fixed order. Phonemic cues were Mandarin syllables (yi, yu, gong, du, wu), with responses in any of the four tones accepted; semantic cues were categories (fruits, cities, vegetables, transport, animals). In both paradigms, every block was followed by a 15-s rest with a fixation cross (+); runs lasted approximately 4 min (A) and 8 min (B).}
\label{fig:1}
\end{figure}

Participants were instructed to name each picture aloud as it appeared on the screen. Each picture was presented for 3 s, with 10 pictures comprising one 30-s task block (Fig.~\ref{fig:1}A). Five task blocks alternated with 15-s rest periods, resulting in a total duration of approximately 4 min \cite{ref30}. Two researchers independently recorded the number of correctly named pictures, with each correct response assigned one point. The records were checked after the task to determine the final score.

\paragraph{Language-fluency task}\label{language-fluency-task}

The task consisted of phonemic and semantic fluency conditions, each comprising five 30-s blocks presented in a fixed order, with phonemic fluency preceding semantic fluency (Fig.~\ref{fig:1}B). During phonemic fluency, participants were instructed to produce as many words or idioms as possible beginning with a specified Mandarin syllable (yi, yu, gong, du or wu), with all four Mandarin tones accepted. During semantic fluency, participants were asked to name as many items as possible from a specified category (fruits, cities, vegetables, means of transportation or animals). Each task block was followed by a 15-s rest period, and the total duration was approximately 8 min. Two researchers independently recorded the number of correct responses, with each correct word assigned one point. The records were checked after the task to determine the final score.

\subsection{Analysis methods}\label{analysis-methods}

\subsubsection{fNIRS signal preprocessing}\label{fnirs-signal-preprocessing}

Resting-state and task-related fNIRS recordings were preprocessed using the same pipeline implemented in NIRS-KIT \cite{ref31}. The preprocessing procedure consisted of first-order linear detrending, motion artefact correction using temporal derivative distribution repair (TDDR) and third-order Butterworth band-pass filtering (0.01--0.08 Hz) applied using zero-phase forward--backward filtering \cite{ref32}. All recordings were processed as continuous runs, with task data filtered before task events were modelled.

\subsubsection{Multi-view fNIRS feature extraction}\label{multi- view-feature-extraction}

The 110 channels were assigned to 22 anatomically defined regions of interest (ROIs) according to their spatial coordinates, and signals from valid channels (those without channel failure or recurrent blockwise signal loss, as assessed before preprocessing) were averaged within each ROI to obtain regional time series (Fig.~\ref{fig:2}, Tables \ref{tab:supp-mni-coordinates} and \ref{tab:supp-roi-assignment} in Appendix).

\begin{figure}[t]
\centering
\includegraphics[width=\textwidth,height=0.65\textheight,keepaspectratio]{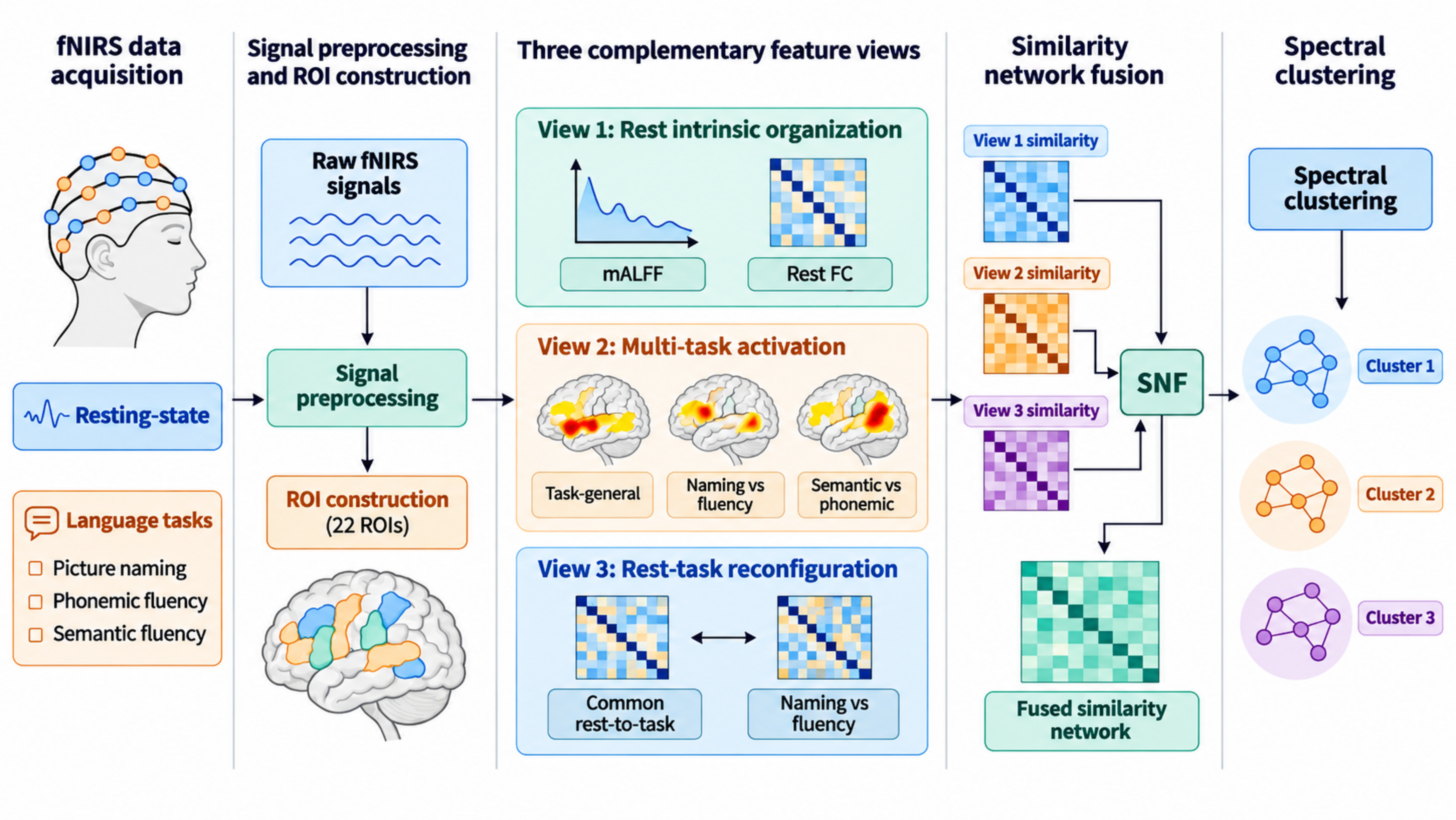}
\caption{\textbf{Workflow of the multi-view fNIRS analysis of post-stroke aphasia for subtyping neurophysiological aphasia.} fNIRS data were acquired at rest and during three language tasks (picture naming, phonemic fluency and semantic fluency), preprocessed and assigned to 22 regions of interest. Three feature views were derived: View 1, resting-state intrinsic organisation, comprising mean amplitude of low-frequency fluctuations (mALFF) and resting-state functional connectivity (FC); View 2, multi-task activation, comprising task-general activation and naming-versus-fluency and semantic-versus-phonemic contrasts; and View 3, rest--task reconfiguration, comprising common rest-to-task connectivity changes and a naming-versus-fluency connectivity contrast. View-specific patient-similarity networks were integrated by similarity network fusion (SNF), and spectral clustering was applied to the fused network to identify three neurophysiological clusters. Abbreviations: fNIRS, functional near-infrared spectroscopy.}
\label{fig:2}
\end{figure}

\textbf{View 1}: Rest intrinsic organization characterised spontaneous regional activity and connectivity over the first 180 s of resting-state recordings. Regional activity was quantified as the amplitude of low-frequency fluctuations normalised by the mean across all ROIs (mALFF; 22 features), whereas Fisher-\emph{z}-transformed Pearson correlations between ROI time series quantified connectivity (231 edges; 22 + 231 = 253 features) \cite{ref33,ref34}.

\textbf{View 2}: Multi-task activation characterized regional responses during naming, phonemic fluency and semantic fluency. First-level general linear models (GLMs) incorporating task timing and a canonical haemodynamic response function were fitted to the continuous task recordings, without further filtering during model fitting \cite{ref35}. Condition-specific coefficients were estimated at the channel level and averaged within each ROI; three predefined contrasts were then computed across conditions, indexing task-general activation, naming-versus-fluency activation and semantic-versus-phonemic activation, each comprising 22 ROI features (3 \(\times\) 22 = 66 features; see Appendix \ref{view-2-multi-task-activation}).

\textbf{View 3}: Rest--task reconfiguration characterised connectivity changes from rest to language tasks. Task-locked responses were first removed by including the canonical haemodynamic response function and its temporal derivative as nuisance regressors. Background connectivity was then estimated separately for naming and fluency from Fisher-\emph{z}-transformed Pearson correlations of residual ROI time series (231 edges each; 231 \(\times\) 2 = 462 features) \cite{ref36}. Two contrasts captured the average rest-to-task change across both tasks and the difference in background connectivity between naming and fluency.

\subsubsection{Similarity network fusion and spectral clustering}\label{similarity-network-fusion-and-spectral-clustering}

Three feature views were integrated using similarity network fusion (SNF) to identify neurophysiological clusters among the 79 patients included in the study \cite{ref37}. Within each view, features were \emph{z}-score standardized, and seven predefined feature families were weighted by the inverse square root of their size to prevent larger families from dominating patient similarities. Pairwise Euclidean distances were converted into patient similarity networks using locally scaled Gaussian kernels \cite{ref38}. SNF then iteratively integrated these equally weighted networks into a single representation of patient similarities.

Spectral clustering was applied to the fused network using a symmetric normalised graph Laplacian \cite{ref39}. Candidate solutions comprising two to six clusters were examined using the eigengap criterion, defined as the difference between consecutive Laplacian eigenvalues. Statistical support for each candidate solution was assessed by permutation testing (see Section \ref{statistical-analysis}). Clinical assessments and behavioural scores were not used for clustering or cluster-number selection. Neighborhood size, kernel, iteration parameters and implementation details are provided in Appendix \ref{supplementary-methods-similarity-network-fusion-and-spectral-clustering}.

\subsubsection{Neurophysiological profiling}\label{neurophysiological-profiling}

Neurophysiological profiles were characterized using the cluster assignments obtained from spectral clustering, with 17 healthy controls as the reference group. Profiling was based on HbO-derived features: 22 regional mALFF values, 66 activation features comprising the 22 ROI-level coefficients for each of the three language tasks (3 \(\times\) 22 = 66) and 18 anatomically grouped functional connectivity summaries computed across three contexts (rest, naming-minus-rest and fluency-minus-rest), three edge groupings and two summary statistics (3 \(\times\) 3 \(\times\) 2 = 18).

Representative features were identified by combining two comparisons: differences from healthy controls and separation of patient clusters. Features meeting both criteria were designated cluster-characteristic, whereas those separating clusters without differing from controls were designated cluster-differentiating; directionally consistent features meeting neither criterion were retained as exploratory descriptions. Changes consistent across all clusters were examined to characterise shared profiles \cite{ref40}. Bootstrap recurrence, the proportion of resamples in which a feature retained its direction of effect, was used to assess feature stability. Because cluster identification and profiling drew on overlapping fNIRS features, between-cluster comparisons were treated as post-clustering descriptions rather than independent validation. Definitions of the anatomically grouped functional-connectivity summaries summaries and full selection criteria are provided in Appendix \ref{supplementary-methods-neurophysiological-profiling}.

\subsubsection{Clinical and behavioural characterization}\label{clinical-and-behavioural-characterisation}

Clinical and behavioural characteristics were examined in all 79 patients. Conventional aphasia subtypes were assigned by clinical judgement according to the WAB (Table~\ref{tab:1}). Clinical language measures comprised WAB total scores (0--415, rather than aphasia quotients) and BDAE severity ratings (1--5) from the initial clinical assessment.

Behavioural performance was quantified as the number of correct responses in each of the three language tasks: picture naming, phonemic fluency and semantic fluency. Clinical scores and subtype distributions were compared across patient clusters, whereas task performance was compared across the three clusters and the healthy control group.

\subsubsection{Statistical analysis}\label{statistical-analysis}

All analyses of cluster characteristics were conducted with cluster assignments held fixed. Candidate solutions (\emph{K} = 2--6) were evaluated using a feature-permutation procedure based on the eigengap statistic; in each of 5,000 permutations the mALFF features were permuted jointly to preserve their within-participant structure, the remaining features were permuted independently and the full pipeline was re-run, with empirical \emph{P} values adjusted across candidates using the Holm method \cite{ref41}.

Differences from healthy controls across the 106 features were assessed using Mann--Whitney U tests, with the signed area under the receiver operating characteristic curve ($\text{AUC}_{\text{signed}}$) as the effect size \cite{ref42}, and differences among clusters using Kruskal--Wallis tests followed by pairwise Mann--Whitney U tests. Benjamini--Hochberg false discovery rate (FDR) correction was applied within prespecified testing families \cite{ref43}, and cluster-differentiating features additionally required an absolute signed AUC of at least 0.35 and a bootstrap recurrence of at least 0.80. Age, post-onset duration and education were compared across clusters using Kruskal--Wallis tests, with Holm correction applied within the three-variable test family and within the family of pairwise comparisons. Clinical and behavioural variables were analysed using the same tests, with Holm correction applied within their respective families. Associations between cluster membership and aphasia subtype were assessed using a fixed-margin Monte Carlo chi-square permutation test, with Cramér's \emph{V} as the effect size \cite{ref44}.

Robustness was assessed using five complementary approaches: patient subsampling (1,000 iterations, each retaining a random 80\% of patients, that is, 64 of 79), variation of the SNF parameters across a 5 \(\times\) 4 grid of neighbourhood size and iteration values (19 configurations in addition to the primary), leave-one-view-out analysis, alternative fusion by equal-weight averaging and leave-one-patient-out clustering (79 runs, each excluding one patient and repeating the full pipeline at fixed \emph{K} = 3) \cite{ref45}. Agreement with the reference partition was quantified using the adjusted Rand index (ARI) and cluster-wise Jaccard indices after Hungarian matching \cite{ref46}. Covariate-adjusted comparisons were restricted to the patient clusters; the connectivity summaries were re-analysed with age as a covariate, and secondarily with age and sex, after rank-based inverse normal transformation of the dependent variable and using heteroscedasticity-consistent standard errors. Parameter values and full sensitivity procedures are provided in Appendix.

\section{Results}\label{results}

\subsection{Identification of neurophysiological clusters}\label{identification-of-neurophysiological-clusters}

\begin{figure}[t]
\centering
\includegraphics[width=\textwidth,height=0.65\textheight,keepaspectratio]{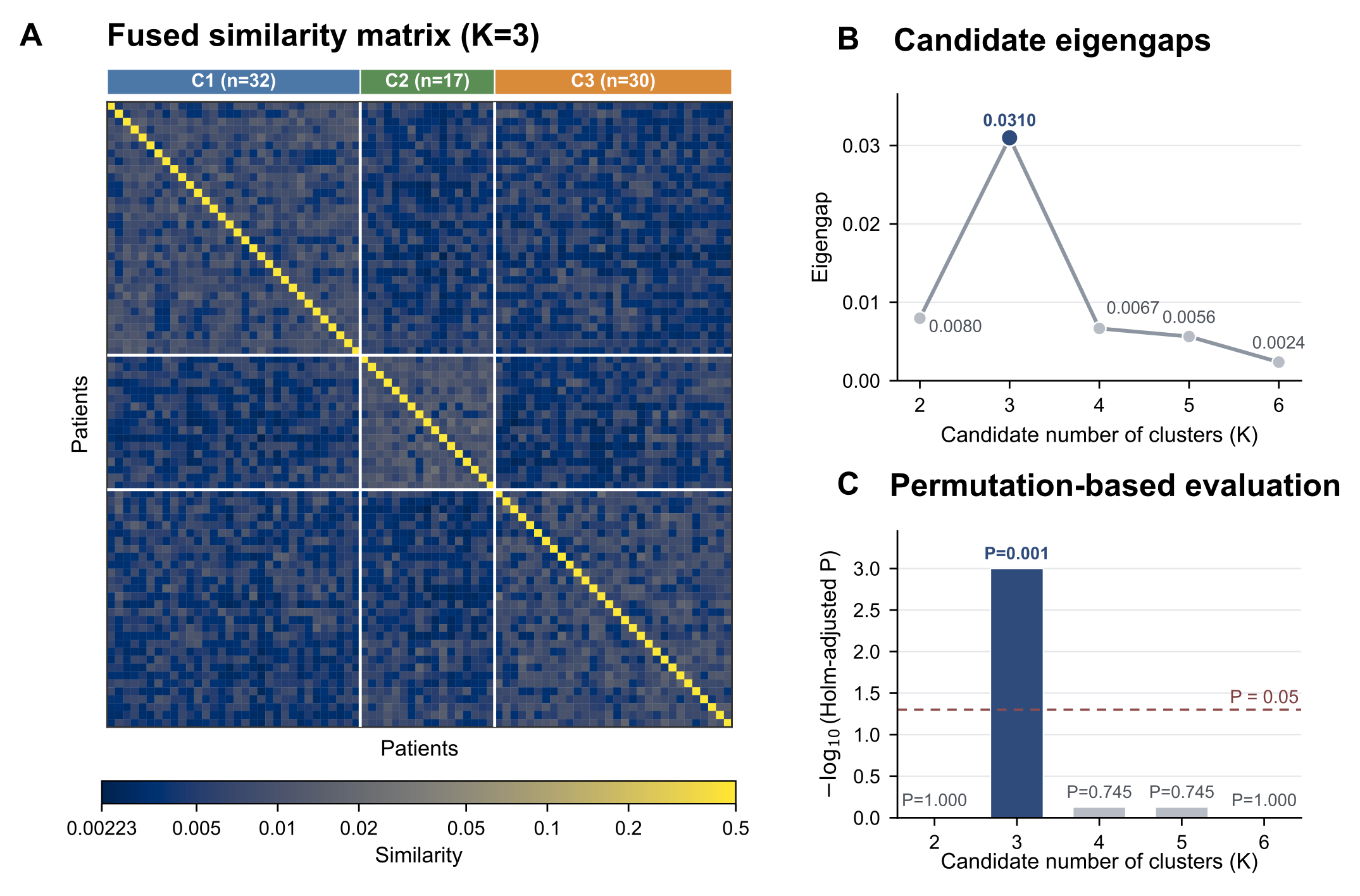}
\caption{\textbf{Identification of neurophysiological clusters of post-stroke aphasia subtypes using multi-view fNIRS.} \textbf{(A)} Fused patient-similarity matrix for the 79 patients, ordered by cluster: C1 (\emph{n} = 32), C2 (\emph{n} = 17) and C3 (\emph{n} = 30), on a logarithmic colour scale. \textbf{(B)} Normalised-Laplacian eigengaps for \emph{K} = 2--6; \emph{K} = 3 gave the largest eigengap (0.0310). \textbf{(C)} Permutation-based evaluation, showing \(-\log_{10}\) Holm-adjusted \emph{P} values from 5,000 feature permutations, with the dashed line marking \emph{P} = 0.05. \emph{K} = 3 was the only solution significant after Holm correction (adjusted \emph{P} = 0.001). Abbreviations: fNIRS, functional near-infrared spectroscopy.}
\label{fig:3}
\end{figure}

Spectral clustering was applied to the fused network, and candidate solutions with two to six clusters were compared. The three-cluster solution showed the largest eigengap (0.0310), compared with 0.0080, 0.0067, 0.0056 and 0.0024 for \emph{K} = 2, 4, 5 and 6, respectively (Fig.~\ref{fig:3}B). In the feature-permutation analysis (5,000 permutations), \emph{K} = 3 was the only candidate that remained significant after Holm correction: the observed eigengap exceeded that of all permuted datasets (empirical \emph{P} \textless{} 0.0002; Holm-adjusted \emph{P} = 0.001). The corresponding adjusted \emph{P} values for \emph{K} = 2, 4, 5 and 6 were 1.000, 0.745, 0.745 and 1.000, respectively (Fig.~\ref{fig:3}C).

The \emph{K} = 3 solution partitioned the 79 patients into three clusters of 32, 17 and 30 patients (C1, C2 and C3; Fig.~\ref{fig:3}A). Demographic and clinical characteristics of these patients, including their conventional WAB-defined aphasia subtypes, are summarized in Table~\ref{tab:2}.

\begin{table}[t]
\centering
\caption{Demographic and clinical characteristics of the clustering cohort}
\label{tab:2}
\footnotesize
\setlength{\tabcolsep}{2.3pt}
\renewcommand{\arraystretch}{1.15}
\begin{tabular}{>{\raggedright\arraybackslash}p{5.05cm}*{4}{>{\centering\arraybackslash}p{2.10cm}}}
\toprule
\textbf{Characteristic} & \textbf{\shortstack{Overall\\(N=79)}} & \textbf{\shortstack{Cluster 1\\(n=32)}} & \textbf{\shortstack{Cluster 2\\(n=17)}} & \textbf{\shortstack{Cluster 3\\(n=30)}} \\
\midrule
Sample size, n & 79 & 32 & 17 & 30 \\
\multicolumn{5}{l}{\textbf{Demographic characteristics}} \\
\addlinespace[2pt]
Age, years; median (Q1--Q3) & \shortstack{62\\(55--69.5)} & \shortstack{64.5\\(58--70)} & \shortstack{69\\(62--74)} & \shortstack{57.5\\(50.25--61.75)} \\
Education, years; median (Q1--Q3) & \shortstack{9\\(6--12)} & \shortstack{9\\(6--12)} & \shortstack{9\\(6--12)} & \shortstack{9\\(6--12)} \\
Post-onset duration, days; median (Q1--Q3) & \shortstack{24\\(16--62.5)} & \shortstack{27\\(19--79.5)} & \shortstack{19\\(14--33)} & \shortstack{23\\(16--83)} \\
Female, n (\%) & 14 (17.7\%) & 6 (18.8\%) & 1 (5.9\%) & 7 (23.3\%) \\
Male, n (\%) & 65 (82.3\%) & 26 (81.2\%) & 16 (94.1\%) & 23 (76.7\%) \\
\multicolumn{5}{l}{\textbf{Stroke type}} \\
\addlinespace[2pt]
Ischaemic stroke, n (\%) & 59 (74.7\%) & 23 (71.9\%) & 14 (82.4\%) & 22 (73.3\%) \\
Haemorrhagic stroke, n (\%) & 13 (16.5\%) & 6 (18.8\%) & 1 (5.9\%) & 6 (20.0\%) \\
Mixed stroke, n (\%) & 7 (8.9\%) & 3 (9.4\%) & 2 (11.8\%) & 2 (6.7\%) \\
\multicolumn{5}{l}{\textbf{Clinical characteristics}} \\
\addlinespace[2pt]
\multicolumn{5}{l}{\textbf{WAB aphasia subtype}} \\
\addlinespace[2pt]
Anomic, n (\%) & 44 (55.7\%) & 17 (53.1\%) & 13 (76.5\%) & 14 (46.7\%) \\
Broca's, n (\%) & 7 (8.9\%) & 4 (12.5\%) & 0 (0.0\%) & 3 (10.0\%) \\
Conduction, n (\%) & 1 (1.3\%) & 0 (0.0\%) & 0 (0.0\%) & 1 (3.3\%) \\
Global, n (\%) & 9 (11.4\%) & 5 (15.6\%) & 1 (5.9\%) & 3 (10.0\%) \\
Transcortical motor, n (\%) & 4 (5.1\%) & 2 (6.2\%) & 0 (0.0\%) & 2 (6.7\%) \\
Transcortical sensory, n (\%) & 3 (3.8\%) & 0 (0.0\%) & 1 (5.9\%) & 2 (6.7\%) \\
Mixed transcortical, n (\%) & 2 (2.5\%) & 0 (0.0\%) & 0 (0.0\%) & 2 (6.7\%) \\
Wernicke's, n (\%) & 9 (11.4\%) & 4 (12.5\%) & 2 (11.8\%) & 3 (10.0\%) \\
\multicolumn{5}{l}{\textbf{Baseline language and behavioural characteristics}} \\
\addlinespace[2pt]
WAB baseline; median (Q1--Q3) & \shortstack{321\\(179.5--376)} & \shortstack{335.5\\(185.75--374.25)} & \shortstack{335\\(290--382)} & \shortstack{293.5\\(148.5--371)} \\
BDAE baseline; median (Q1--Q3) & \shortstack{3\\(2--3)} & \shortstack{3\\(2--3)} & \shortstack{3\\(2--3)} & \shortstack{2.5\\(2--3.75)} \\
Naming; median (Q1--Q3) & \shortstack{29\\(8--40.5)} & \shortstack{34\\(6--41)} & \shortstack{31\\(17--37)} & \shortstack{22\\(7.25--38.75)} \\
Phonemic fluency; median (Q1--Q3) & \shortstack{6\\(1.5--10)} & \shortstack{6\\(0.75--10.25)} & \shortstack{6\\(2--7)} & \shortstack{4.5\\(1.25--9.75)} \\
Semantic fluency; median (Q1--Q3) & \shortstack{20\\(4--29)} & \shortstack{19\\(2.25--32.25)} & \shortstack{22\\(12--30)} & \shortstack{14.5\\(2--26.75)} \\
\bottomrule
\end{tabular}
\vspace{0.6em}
\begin{minipage}{\textwidth}
\footnotesize
Values are presented as median (Q1-Q3) or n (\%), as indicated.

The table includes 79 patients with matched demographic and clinical data; cluster sizes were 32, 17, and 30. Percentages are calculated using the available sample within each column.

Post-onset duration is reported in days. WAB baseline is the supplied 0-415 score and not WAB Aphasia Quotient. BDAE baseline is the recorded 1-5 score.

BDAE, Boston Diagnostic Aphasia Examination; WAB, Western Aphasia Battery.
\end{minipage}
\end{table}

\subsection{Neurophysiological profiles of the three clusters}\label{neurophysiological-profiles-of-the-three-clusters}

\begin{figure}[t]
\centering
\includegraphics[width=\textwidth,height=0.65\textheight,keepaspectratio]{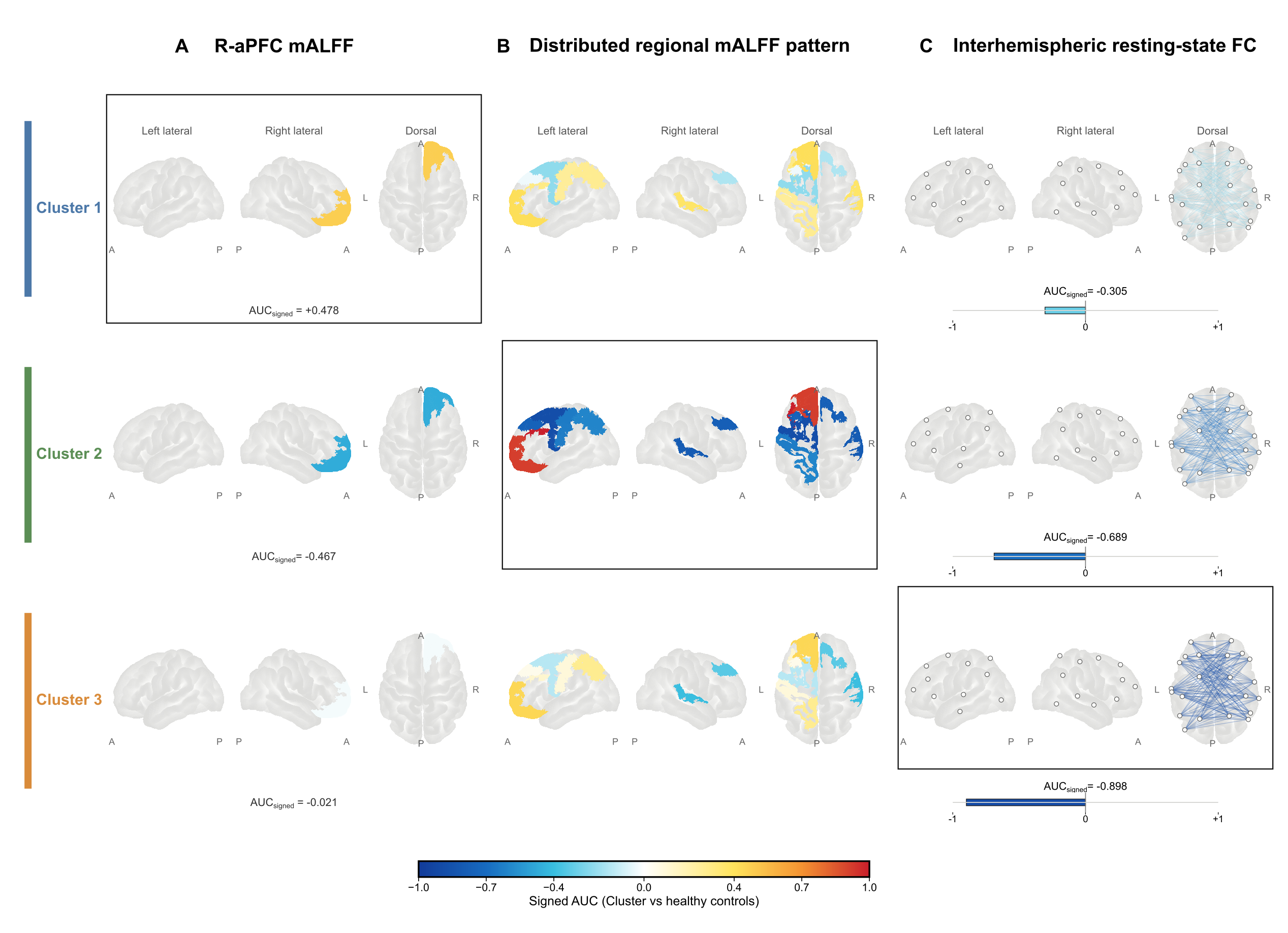}
\caption{\textbf{Distinct neurophysiological profiles of three post-stroke aphasia subtype clusters.} Rows show C1 (\emph{n} = 32), C2 (\emph{n} = 17) and C3 (\emph{n} = 30), each compared with 17 healthy controls. Colour encodes cluster-versus-control Signed AUC ($\text{AUC}_\text{signed}$), from \(-1\) to +1, with positive values indicating higher measurements than in controls. \textbf{(A)} Mean amplitude of low-frequency fluctuations (mALFF) in right anterior prefrontal cortex. \textbf{(B)} Distributed mALFF across eight regions of frontal, sensorimotor, parietal and temporal cortex. \textbf{(C)} Mean signed interhemispheric resting-state functional connectivity across 121 left--right ROI pairs; line colour reflects the cluster-level summary effect, not edge-specific inference. Thin black frames mark prespecified representative features. Brain maps use the FreeSurfer fsaverage5 surface in lateral and dorsal views. Abbreviations: $\text{AUC}_\text{signed}$, signed area under the receiver operating characteristic curve; ROI, region of interest.}
\label{fig:4}
\end{figure}

Of the 22 regional mALFF features, 19 showed between-cluster differences after FDR correction. Right anterior prefrontal mALFF was highest in C1, lowest in C2 and at the level of controls in C3, and all pairwise cluster contrasts survived correction (\emph{q} = 0.0009--0.0419). Beyond this shared axis the clusters diverged: C1 showed elevated left superior temporal mALFF, C2 a distributed pattern of frontal, sensorimotor and parietal alterations with complete separation from controls in the left anterior and left dorsolateral prefrontal cortex ($\text{AUC}_{\text{signed}}$ = +1.000), and C3 focal reductions in right inferior frontal and right posterior occipitotemporal cortex (Fig.~\ref{fig:4}A and B).

All six resting-state functional-connectivity summaries based on anatomical edge groupings differed between clusters (\emph{q} = \(2.45 \times 10^{-6}\)--\(6.82 \times 10^{-4}\)). The reduction was most pronounced in C3, which had the lowest mean signed interhemispheric FC (0.079 versus 0.390 in controls; $\text{AUC}_{\text{signed}}$ = \(-0.898\), \emph{q} = \(8.4 \times 10^{-5}\)) and differed from both other clusters. C2 showed a moderate reduction (0.169; $\text{AUC}_{\text{signed}}$ = \(-0.689\), \emph{q} = 0.011), whereas C1 did not differ from controls (Fig.~\ref{fig:4}C).

Task-related activation and rest--task reconfiguration contributed little to this separation. Only one of the 66 activation features showed a between-cluster effect (phonemic-fluency \emph{\(\beta\)} in right anterior prefrontal cortex, \emph{q} = 0.033), with no surviving pairwise contrast, and none of the 12 reconfiguration summaries showed any effect. All 106 features, including those that did not reach significance, are shown in Fig.~\ref{fig:supp-profiles} in Appendix Section.

Together, these results indicate that the clusters were distinguished by different combinations of regional activity and connectivity alterations, superimposed on a common gradient of right anterior prefrontal mALFF.

\subsection{Clinical and behavioural characterisation}\label{clinical-and-behavioural-characterisation-1}

Age differed across the three clusters (median 64.5, 69 and 57.5 years in C1, C2 and C3; \emph{H}(2) = 14.38, Holm-adjusted \emph{P} = 0.002), mainly because C3 was younger than C2 (Holm-adjusted \emph{P} = 0.006). Due to that the age can potentially affect resting-state FC, the between-cluster FC comparisons were repeated with age as a covariate. The cluster effect persisted (omnibus \emph{q} = \(1.2 \times 10^{-5}\)) and the C1--C3 contrast remained significant (\emph{q} = \(6.6 \times 10^{-6}\)), whereas the C2--C3 contrast kept its direction but fell just above the adjusted threshold (uncorrected \emph{P} = 0.007; \emph{q} = 0.053). This imbalance does not account for the reduction observed in C3, which was younger than the control group (median 57.5 years; control mean 61 \(\pm\) 8 years), so that age-related decline would predict the opposite direction of effect.

\begin{figure}[t]
\centering
\includegraphics[width=\textwidth,height=0.65\textheight,keepaspectratio]{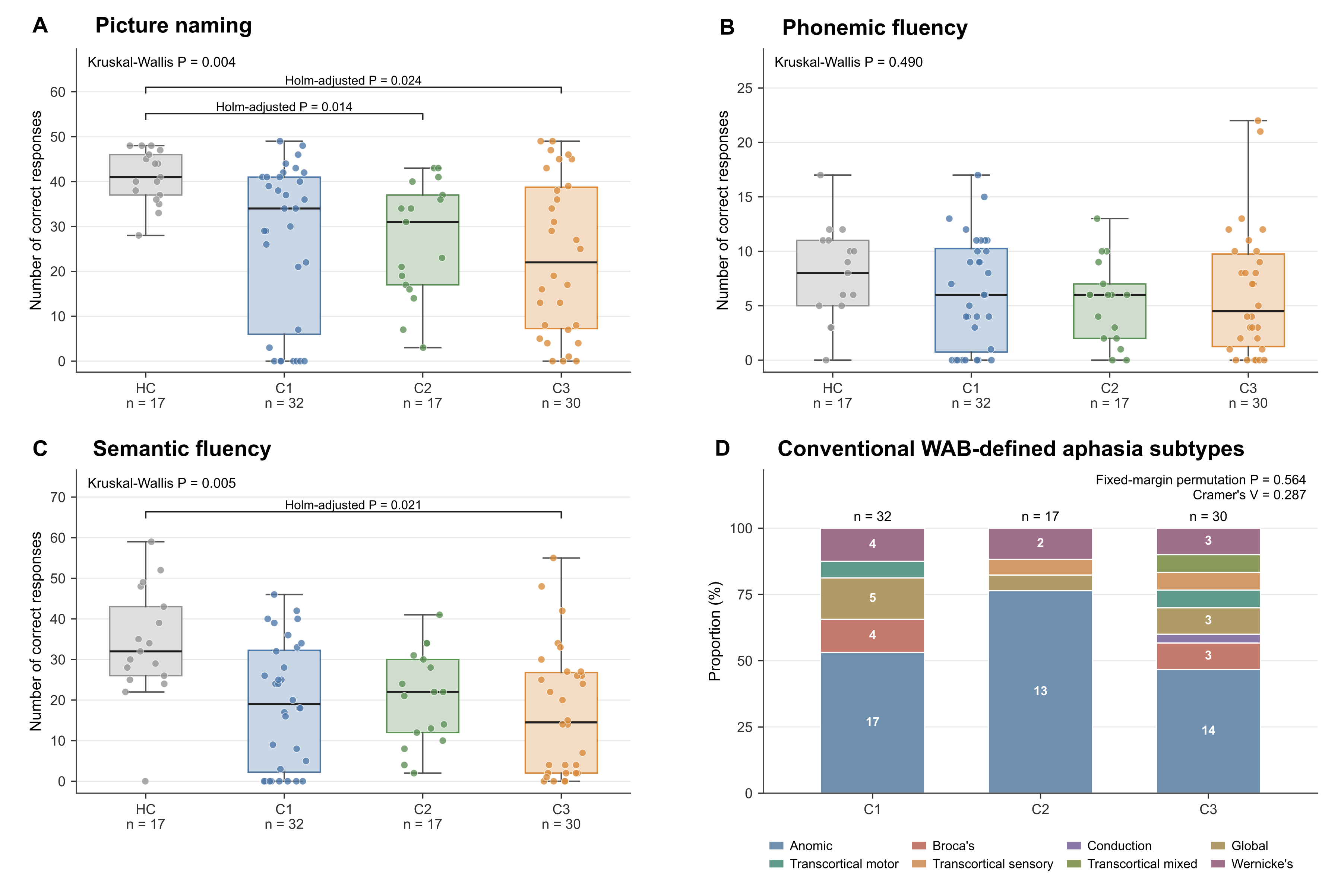}
\caption{\textbf{Behavioural performance and conventional aphasia subtype composition across healthy controls and the identified three subtypes.} \textbf{(A--C)} Correct responses in picture naming \textbf{(A)}, phonemic fluency (B) and semantic fluency (C) for healthy controls (HC; \emph{n} = 17) and clusters C1 (\emph{n} = 32), C2 (\emph{n} = 17) and C3 (\emph{n} = 30). Boxes span the interquartile range, centre lines the median, whiskers 1.5 times the interquartile range, and individual observations are overlaid. \emph{P} values above each panel derive from Kruskal--Wallis tests. Brackets mark cluster-versus-HC comparisons surviving Holm correction across 18 tests; no comparison between patient clusters survived. \textbf{(D)} Composition of eight WAB-defined aphasia subtypes within each cluster, in fixed order and with observed counts in each segment; the association with cluster membership was assessed by a fixed-margin Monte Carlo \(\chi^2\) permutation test, with Cramér's \emph{V} as the effect size. Abbreviations: WAB, Western Aphasia Battery.}
\label{fig:5}
\end{figure}

Conventional clinical measures did not distinguish the three clusters. Median WAB scores were 335.5, 335 and 293.5 in C1, C2 and C3, respectively (\emph{H}(2) = 0.99, \emph{P} = 0.610), and median BDAE scores were 3, 3 and 2.5 (\emph{H}(2) = 0.16, \emph{P} = 0.925); neither differed across clusters after Holm correction. Anomic aphasia was the most frequent subtype in all three clusters, ranging from 46.7\% in C3 to 76.5\% in C2 (see Table~\ref{tab:2}), and cluster membership provided no evidence of an association with conventional subtype (fixed-margin permutation \emph{P} = 0.564; Cramér's \emph{V} = 0.287; Fig.~\ref{fig:5}D).

Behavioural performance showed descriptive differences across clusters (Fig.~\ref{fig:5}A--C). Median picture naming was highest in C1 (34 correct responses) and median semantic fluency highest in C2 (22), whereas C3 had the lowest median on all three tasks (22, 4.5 and 14.5 for picture naming, phonemic fluency and semantic fluency, respectively). Across the four groups, picture naming (\emph{P} = 0.004) and semantic fluency (\emph{P} = 0.005) differed, whereas phonemic fluency did not (\emph{P} = 0.490). Compared with controls, picture naming was lower in C2 and C3 (Holm-adjusted \emph{P} = 0.014 and 0.024, respectively) and semantic fluency lower in C3 (Holm-adjusted \emph{P} = 0.021). Within the patient cohort, however, no behavioural measure differed across clusters (all \emph{P} \(\geq\) 0.605), and no pairwise cluster comparison survived correction.

Considered alongside their neurophysiological profiles, these results indicate that the clusters did not recapitulate conventional aphasia classification or a single gradient of language impairment.

\subsection{Robustness and sensitivity analyses}\label{robustness-and-sensitivity-analyses}

\begin{figure}[t]
\centering
\includegraphics[width=\textwidth,height=0.65\textheight,keepaspectratio]{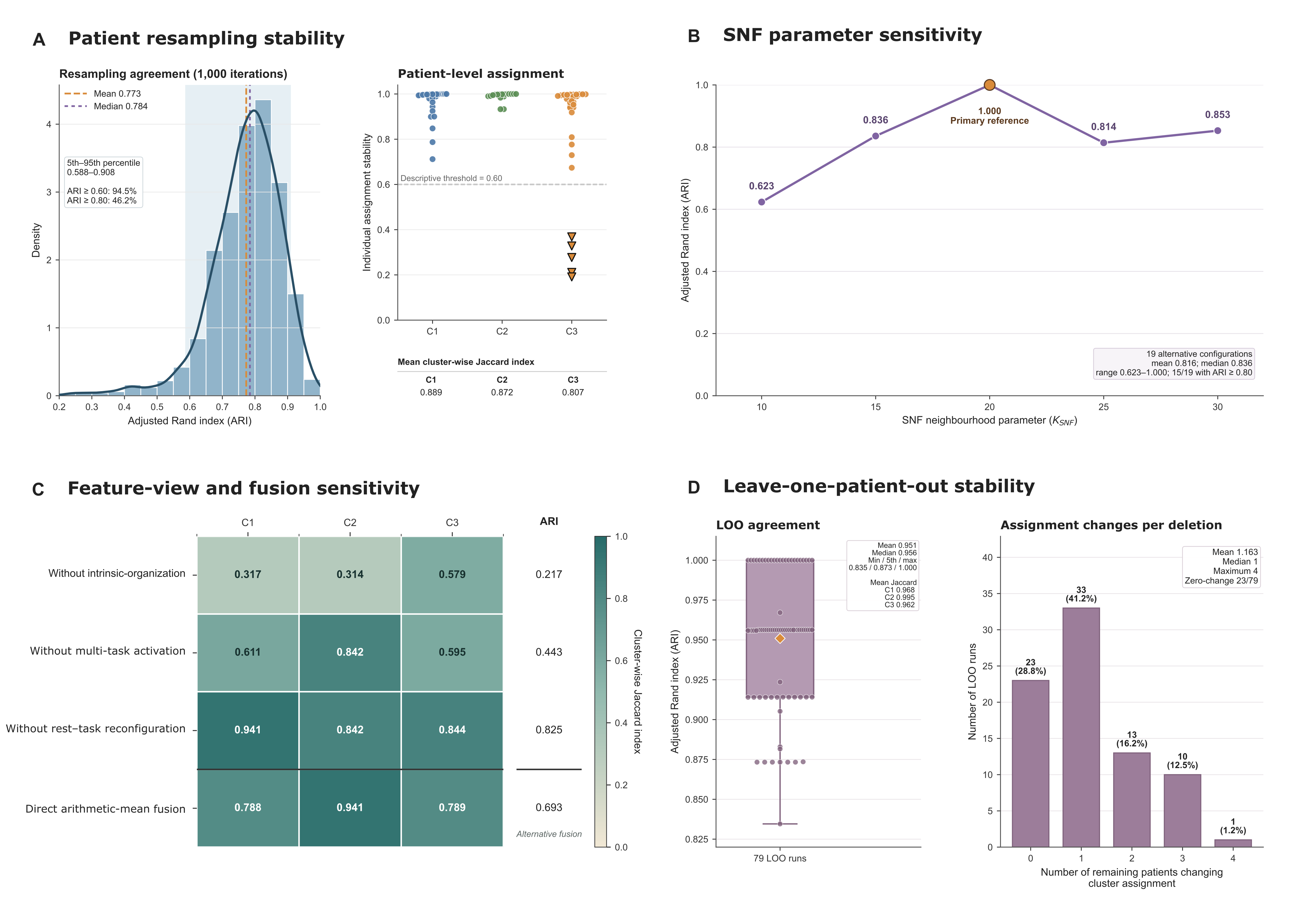}
\caption{\textbf{Internal robustness of the \emph{K} = 3 clustering solution.} \textbf{(A)} Resampling stability across 1,000 iterations retaining 64 of 79 patients. The histogram and density curve show agreement with the reference partition (ARI; mean 0.773, median 0.784, 5th--95th percentile 0.588--0.908). The adjacent strip plot shows individual assignment stability by cluster; triangles mark the five patients below the 0.60 threshold, all in C3. \textbf{(B)} Sensitivity analysis of the SNF's neighbourhood parameter ($\emph{K}_{SNF}$). The orange point marks the primary configuration, whose ARI of 1.000 is a self-comparison rather than independent validation; across 19 alternative configurations, mean ARI was 0.816 and 15 reached 0.80. \textbf{(C)} Cluster-wise Jaccard indices after leave-one-view-out analysis and arithmetic-mean fusion; the right-hand column gives the overall ARI, which is not encoded by the colour scale. \textbf{(D)} Leave-one-patient-out stability across 79 runs at fixed \emph{K} = 3; the histogram counts runs by the number of remaining patients reassigned. Abbreviations: ARI, adjusted Rand
index; SNF, similarity network fusion.}
\label{fig:6}
\end{figure}

Agreement with the reference partition was high across perturbation conditions (Fig.~\ref{fig:6}). Across 1,000 subsampling iterations, each retaining 64 of the 79 patients (80\%), the mean ARI was 0.773 (median, 0.784; 5th--95th percentile, 0.588--0.908). Mean cluster-wise Jaccard indices were 0.889, 0.872 and 0.807 for C1, C2 and C3, and five patients in C3 had individual assignment stability below 0.60 (Fig.~\ref{fig:6}A). Across the 19 alternative SNF parameter configurations, mean ARI was 0.816 and 15 configurations reached 0.80; changing the iteration parameter did not alter discrete assignments at any neighbourhood size (Fig.~\ref{fig:6}B). Leave-one-patient-out analysis across 79 runs gave a mean ARI of 0.951 (median, 0.956; minimum, 0.835), with a mean of 1.16 remaining patients reassigned per run, at most four changes in any run and no change in 23 of 79 runs (Fig.~\ref{fig:6}D).

Agreement depended on which views were retained (Fig.~\ref{fig:6}C). Removing the intrinsic-organisation view produced the lowest agreement (ARI, 0.217; Jaccard, 0.314--0.579), followed by removal of the multi-task activation view (ARI, 0.443; Jaccard, 0.595--0.842), whereas removing the rest--task reconfiguration view retained greater agreement (ARI, 0.825; Jaccard, 0.842--0.941). Equal-weight arithmetic-mean fusion yielded an ARI of 0.693.

Neurophysiological profiles were directionally stable across these analyses. Bootstrap direction consistency ranged from 91.1\% to 100\% and no signed AUC reversed direction in leave-one-patient-out profiling. Age adjustment, restricted to between-cluster comparisons, preserved 10 of the 12 feature--cluster combinations that met the prespecified criteria for representative profiling (nine cluster-characteristic and three cluster-differentiating).

\section{Discussion}\label{discussion}

We identified three candidate neurophysiological subgroups of post-stroke aphasia by integrating resting-state and task-based fNIRS data across three feature views. The profiles were driven primarily by distinct patterns of regional intrinsic activity and resting-state functional connectivity; task-related activation and connectivity reconfiguration contributed comparatively little to cluster separation. The clusters were not significantly associated with conventional WAB-defined aphasia subtypes or baseline clinical language scores, suggesting that fNIRS-derived profiling may capture aspects of functional brain organization not fully represented by syndrome-based classification. These findings indicate that multi-view neurophysiological profiling provides a complementary perspective on patient heterogeneity that is not readily captured by syndrome-based classification or clinical language scores alone.

\subsection{Neurophysiological heterogeneity in post-stroke aphasia}\label{neurophysiological-heterogeneity-in-post-stroke-aphasia}

The three clusters showed different combinations of regional intrinsic activity and functional connectivity rather than a uniform pattern of functional alteration. C1 was characterized by relatively higher right anterior prefrontal mALFF than C2 and C3, whereas C2 showed a more distributed pattern of elevated regional mALFF involving frontal, sensorimotor and parietal regions, including complete separation from controls in left prefrontal cortex. C3, in contrast, showed the most pronounced reduction in interhemispheric resting-state functional connectivity (FC), with mean signed interhemispheric FC of 0.079 compared with 0.390 in healthy controls. C2 also showed reduced interhemispheric connectivity relative to controls, although this reduction was less pronounced and did not persist consistently after age adjustment. These findings are consistent with previous evidence that post-stroke aphasia is associated with heterogeneous patterns of residual language-network activity and connectivity, both within and beyond canonical language regions \cite{ref15,ref16,ref17,ref18,ref19,ref20,ref47}.

Regional activity differences should not, however, be interpreted as direct evidence of compensatory recruitment. Meta-analytic work indicates that the neural resources engaged after stroke vary across language domains, with language production and comprehension drawing on partly distinct left- and right-hemisphere regions \cite{ref48}. A recent functional imaging study further showed that activity in left and right posterior temporal regions can make independent contributions to language outcomes, with no general support found for right frontal recruitment as a compensatory mechanism \cite{ref49}. Thus, the relatively higher right anterior prefrontal mALFF observed in C1 is best regarded as a distinguishing feature of its intrinsic profile, rather than evidence of a specific compensatory mechanism. The pronounced reduction in interhemispheric connectivity observed in C3 may therefore be of particular interest in this context, given evidence linking language-network connectivity to language improvement after stroke \cite{ref50}, although its prognostic implications remain to be established. The distributed pattern of elevated mALFF observed in C2, spanning frontal, sensorimotor and parietal regions, is less readily mapped onto established frameworks and may reflect a more complex reorganisation that warrants further investigation.

\subsection{Relationship to conventional clinical classification}\label{relationship-to-conventional-clinical-classification}

The neurophysiological clusters did not map directly onto conventional clinical classifications. Cluster membership was not significantly associated with WAB-defined aphasia subtype (\emph{P} = 0.564; Cramér's \emph{V} = 0.287, indicating a moderate association that did not reach significance), and baseline WAB and BDAE scores did not differ significantly across clusters after correction. Although subtype proportions and task scores showed descriptive variation, these patterns did not establish statistically distinct clinical or behavioural profiles. Compared with healthy controls, picture-naming performance was lower in C2 and C3, while semantic fluency was lower in C3; however, these cluster-versus-control differences do not indicate significant behavioural separation among the patient clusters. The observed age difference across clusters --- particularly the younger median age of C3 --- was addressed by repeating the FC comparisons with age as a covariate; the cluster effect persisted, and because C3 was younger than the control group, age-related decline would predict the opposite direction of connectivity change.

The absence of a direct correspondence between neurophysiological clusters and conventional clinical classifications is consistent with evidence that post-stroke aphasia involves multidimensional language components and recovery trajectories, which may have distinct neural correlates \cite{ref12,ref14,ref51}. The present findings extend this perspective by suggesting that neurophysiological profiles may offer a complementary level of patient characterisation that is not readily inferred from syndrome labels or baseline clinical scores alone, and should be viewed as complementary to, rather than replacements for, conventional clinical assessment. This complementary role may be clinically relevant: resting-state network connectivity has been associated with responsiveness to language therapy \cite{ref52}, and changes in structural language-network topology have been related to naming recovery following combined scalp acupuncture and speech-language therapy \cite{ref53}. Whether the neurophysiological clusters identified here relate to treatment response remains to be tested.

\subsection{Implications of multi-view neurophysiological profiling}\label{implications-of-multi-view-neurophysiological-profiling}

The contribution of three feature views to the integrated clustering solution was unequal. Removing the resting-state intrinsic-organisation view produced the greatest change in cluster assignments (ARI = 0.217), whereas removing the rest--task reconfiguration view retained much greater agreement with the full solution; the multi-task activation view occupied an intermediate position. These findings indicate that intrinsic organisation and task-related activation both contributed to the identified partition, whereas the reconfiguration view had a comparatively smaller influence on its overall structure. The ablation results reflect the dependence of the integrated partition on each view, rather than the standalone quality or biological importance of any single feature set. The comparatively limited contribution of the reconfiguration view may partly reflect overlap between task-background connectivity and resting-state connectivity structure, particularly given the restricted variability in task-related neural responses observed in a clinically impaired sample; whether reconfiguration features carry additional discriminative value in cohorts with greater response variability remains to be examined.

Previous fNIRS studies in post-stroke aphasia have examined task-evoked activation and resting-state connectivity, with fewer studies combining these measurements within the same cohort \cite{ref23,ref24,ref25,ref26,ref27}. By integrating intrinsic activity and connectivity, multi-task activation, and rest--task connectivity reconfiguration within a data-driven unsupervised framework, the present approach brings these complementary measurements into a common patient-level representation. Similarity network fusion is particularly suited to this integration, as it preserves view-specific patient-similarity structures rather than assuming a common distance metric across feature types \cite{ref37}. This approach extends a strategy applied in EEG-based subtyping of amyotrophic lateral sclerosis, where data-driven clustering similarly revealed neurophysiological subphenotypes not captured by clinical staging \cite{ref40}. The present findings suggest that the value of multi-view profiling lies not in assuming that all views contribute equally, but in combining partially overlapping information that may be missed when each measurement context is considered in isolation.

\subsection{Limitations and future directions}\label{limitations-and-future-directions}

This study is limited by its single-center design and the absence of independent external validation. Although internal resampling and parameter-sensitivity analyses demonstrated overall stability of the reference partition, view dependence in the leave-one-view-out analysis indicates that the solution relies substantially on the resting-state intrinsic-organization view, and the current partition should therefore be regarded as a candidate stratification rather than a definitive classification. Replication in independent cohorts is needed to establish its reproducibility and generalizability in future.

Several additional characteristics of the study design and sample warrant consideration. Lesion location and volume data were not available for inclusion in the current analyses; because all patients had left-hemisphere lesions, the neurophysiological profiles --- and in particular the interhemispheric connectivity reduction observed in C3 --- may partly reflect variation in lesion extent or distribution rather than purely functional organization, and future studies should integrate structural lesion data to address this. The healthy control group was small (\emph{n} = 17) and was not individually matched to patients by age or other demographic characteristics; the covariate analyses reported here partially address the age imbalance but do not substitute for a matched comparison group. The absence of a significant association between cluster membership and conventional aphasia subtype should also be interpreted with caution, given the sparse distribution of several subtypes in a single-centre cohort of this size; the moderate observed effect size (Cramér's \emph{V} = 0.287) does not exclude a clinically meaningful relationship that a larger or more heterogeneous sample might detect. Finally, task performance was quantified using correct-response counts rather than standardised assessments, which may not fully capture the breadth of language impairment.

The cross-sectional design does not establish whether the identified profiles remain stable across the recovery trajectory or carry prognostic value for language outcomes; longitudinal studies with repeated neurophysiological assessments are needed to address this. More broadly, integrating neurophysiological profiling with structural lesion mapping, treatment response data and standardised longitudinal measures will be necessary to evaluate whether these candidate subgroups can support clinical stratification in aphasia rehabilitation.

\section{Conclusion}\label{conclusion}

Integrating resting-state and task-based fNIRS data across multiple feature views, this study identified three candidate neurophysiological subgroups of post-stroke aphasia, characterized by distinct patterns of intrinsic cortical activity and resting-state functional connectivity. These profiles captured aspects of functional brain organization that were not reflected in conventional syndrome labels or baseline clinical language scores, suggesting that neurophysiological variation in post-stroke aphasia extends beyond the dimensions currently indexed by clinical assessment. The findings suggest that portable multi-condition fNIRS, combined with data-driven integration, can provide a complementary neurophysiological perspective on patient heterogeneity that is not readily accessible through conventional clinical characterization. Independent replication and longitudinal follow-up are needed to establish the reproducibility, temporal stability and clinical utility of these candidate profiles; if validated, neurophysiological subtyping may ultimately inform more targeted approaches to rehabilitation in post-stroke aphasia.

\section{Data availability}\label{data-availability}

The data presented in this study can be made available upon request from the corresponding authors. However, due to privacy and ethical concerns, the data cannot be publicly disclosed. The code used for data preprocessing, feature extraction, clustering, and statistical analyses will be publicly available on GitHub at https://github.com/jtggz/fNIRS-subtypingaphasia.

\section{Funding}\label{funding}
This work was supported by the National Natural Science Foundation of China (No.82575187), the Fujian Province Natural Science Foundation-Young Investigator Project (No. 2024J010033), The Hong Kong Polytechnic University Departmental Collaborative Research Fund (Project ID: P0056428), an internal grant from The Hong Kong Polytechnic University (Project ID: P0048377), The Hong Kong Polytechnic University Collaborative Research with World-leading Research Groups Fund (Project ID: P0058097) and Research Grants Council Collaborative Research Fund (Project ID: C5033-24G).

\section{Competing interests}\label{competing-interests}

The authors report no competing interests.

\bibliographystyle{unsrt}
\bibliography{references}

\setcounter{equation}{0}
\renewcommand{\theequation}{A\arabic{equation}}
\setcounter{table}{0}
\renewcommand{\thetable}{S\arabic{table}}
\setcounter{figure}{0}
\renewcommand{\thefigure}{S\arabic{figure}}

\section{Appendix}\label{sec:supplementary-methods}

\subsection{Multi-view feature extraction}\label{supplementary-methods-multi-view-feature-extraction}

\subsubsection{ROI definition and
aggregation}\label{roi-definition-and-aggregation}

The 110 fNIRS measurement channels were assigned to 22 anatomically defined ROIs according to their spatial coordinates and the predefined channel-to-ROI mapping. Each channel belonged to one ROI. Channel coordinates and ROI assignments are provided in Supplementary Tables \ref{tab:supp-mni-coordinates} and \ref{tab:supp-roi-assignment}, respectively.

For time-series-based analyses, the ROI signal was calculated as thearithmetic mean of valid constituent channels:

\begin{equation}
\label{eq:supp-1}
\textbf{x}_{j}(t) = \frac{1}{n_{j}}\sum_{c \in V_{j}}{\textbf{x}_{c}(t)}
\end{equation}
where \(V_{j}\) denotes the set of valid channels within ROI \(j\), and
\(n_{j}\) is the number of valid channels.

For task activation, GLM coefficients were first estimated at the channel level and subsequently averaged within each ROI:

\begin{equation}
\label{eq:supp-2}
\beta_{j}^{q} = \frac{1}{n_{j}}\sum_{c \in V_{j}}\beta_{c}^{q}
\end{equation}
where \(q\) denotes the naming, phonemic fluency or semantic fluency condition. ROIs without valid channels were not imputed.

\subsubsection{View 1: Rest intrinsic organisation}\label{view-1-rest-intrinsic-organisation}

The first continuous 180 s of resting-state recordings were used to characterize regional spontaneous activity and intrinsic functional connectivity.

Regional activity was quantified using the amplitude of low-frequency fluctuations (ALFF). For each participant, ALFF values were normalized by the mean across all 22 ROIs:

\begin{equation}
\label{eq:supp-3}
\mathrm{mALFF}_{j} = \frac{\mathrm{ALFF}_{j}}{\frac{1}{22}\sum_{k = 1}^{22}\mathrm{ALFF}_{k}}
\end{equation}
This yielded 22 regional activity features.

Resting-state functional connectivity was estimated using Pearson correlation coefficients between ROI time series, followed by Fisher's \emph{z}-transformation:

\begin{equation}
\label{eq:supp-4}
\begin{aligned}
r_{ij}^{R} & = \operatorname{corr}\left( \textbf{x}_{i},\textbf{x}_{j} \right) \\
z_{ij}^{R} & = \operatorname{atanh}\left( r_{ij}^{R} \right)
\end{aligned}
\end{equation}
The upper triangle of the connectivity matrix, excluding the diagonal, provided 231 unique connections. Together, 22 mALFF and 231 connectivity features formed the 253-dimensional intrinsic representation.

\subsubsection{View 2: Multi-task activation}\label{view-2-multi-task-activation}

Task-related activation was estimated using first-level general linear models fitted by ordinary least squares. Task regressors were constructed over each continuous recording, convolved with the canonical haemodynamic response function and filtered consistently with the preprocessed signals. An unfiltered intercept was included, and the signals were not filtered a second time during model fitting.

Naming was modelled as a single condition. Phonemic and semantic fluency were modelled jointly within the continuous fluency recording. Channel-level GLM coefficients were averaged within ROIs, providing activation estimates for the three conditions.

Let \(N_{j}\), \(P_{j}\) and \(S_{j}\) represent the naming, phonemic fluency and semantic fluency coefficients for ROI \(j\), respectively.
Three predefined measures were calculated:
\begin{equation}
\label{eq:supp-5}
\begin{aligned}
G_{j}^{Act} & = \frac{N_{j} + P_{j} + S_{j}}{3} \\
D_{NF,j}^{Act} & = N_{j} - \frac{P_{j} + S_{j}}{2} \\
D_{SP,j}^{Act} & = S_{j} - P_{j}
\end{aligned}
\end{equation}
These measures represented task-general activation, naming-versus-fluency activation and semantic-versus-phonemic activation, respectively. Each contributed 22 ROI features, yielding a 66-dimensional representation.

\subsubsection{View 3: Rest--task reconfiguration}\label{view-3-resttask-reconfiguration}

Task-background functional connectivity was estimated separately for naming and fluency. At the ROI level, task-locked haemodynamic responses were modelled using the canonical haemodynamic response function and its temporal derivative. The temporal derivative served as an additional nuisance regressor for residualisation. Pearson correlations between residual ROI time series were Fisher \emph{z}-transformed to obtain 231 background connectivity values for each task context.

The fluency recording was treated as a single task context for connectivity analysis; phonemic-only and semantic-only connectivity matrices were not constructed.

Let \(z_{ij}^{R}\), \(z_{ij}^{N}\) and \(z_{ij}^{F}\) denote the Fisher-\emph{z} connectivity values for rest, naming and fluency, respectively. Edge-wise rest-to-task changes were defined as:

\begin{equation}
\label{eq:supp-6}
\begin{aligned}
{\Delta N}_{ij} & = z_{ij}^{N} - z_{ij}^{R} \\
{\Delta F}_{ij} & = z_{ij}^{F} - z_{ij}^{R}
\end{aligned}
\end{equation}
Two complementary connectivity measures were then constructed:

\begin{equation}
\label{eq:supp-7}
\begin{aligned}
G_{ij}^{RT} & = \frac{{\Delta N}_{ij} + {\Delta F}_{ij}}{2} \\
D_{NF,ij}^{FC} & = {\Delta N}_{ij} - {\Delta F}_{ij}
\end{aligned}
\end{equation}
The first measure captured the average rest-to-task connectivity change across naming and fluency, whereas the second represented their difference in task-background connectivity. Both retained the direction of Fisher-\emph{z} connectivity differences. Each measure comprised 231 connections, yielding 462 features.

The three views together provided 781 HbO-derived features. Feature standardisation, patient similarity construction and multi-view fusion were performed in subsequent analysis steps.

\subsection{Similarity network fusion and spectral clustering}\label{supplementary-methods-similarity-network-fusion-and-spectral-clustering}

\subsubsection{Feature standardisation and patient similarity networks}\label{feature-standardisation-and-patient-similarity-networks}

The three HbO feature matrices were aligned across the 79 patients with complete data. Each feature was standardized across patients using its sample mean and standard deviation. To balance the contributions of feature families with different dimensionalities, standardized features were further weighted by the inverse square root of their family size.

Seven predefined feature families were used: mALFF (22 features), resting-state FC (231), task-general activation (22), naming-versus-fluency activation (22), semantic-versus-phonemic activation (22), common rest-to-task reconfiguration (231) and naming-versus-fluency background FC differences (231).

Within each view, Euclidean distances were calculated between patients using the standardized and family-weighted features. Distances were converted into affinity matrices using locally scaled Gaussian density kernels, following an implementation compatible with SNFtool version 2.3.1. The affinity construction used 20 nearest neighbours and a kernel parameter of 0.5, producing three 79 \(\times\) 79 patient similarity matrices.

\subsubsection{Similarity network fusion (SFN)}\label{similarity-network-fusion}

The three similarity matrices were integrated using an SNF implementation compatible with SNFtool version 2.3.1. Each view contributed equally to the fusion process. During iterative network updating, patient-neighbourhood information was exchanged across views to integrate their complementary similarity structures.

The number of neighbours was set to 20, and fusion was performed for 20 iterations. The resulting fused similarity matrix was used directly for spectral clustering without additional network enhancement.

\subsubsection{Spectral clustering}\label{spectral-clustering}

Spectral clustering was performed on the fused patient similarity matrix using an implementation compatible with the SNFtool spectral clustering procedure. A symmetric normalized graph Laplacian was constructed:

\begin{equation}
\label{eq:supp-8}
\textbf{L} = \textbf{I} - \textbf{D}^{- \frac{1}{2}}\textbf{WD}^{- \frac{1}{2}}
\end{equation}
where \(\textbf{W}\) is the fused similarity matrix and \(\textbf{D}\) is the diagonal matrix of node degrees.

Candidate cluster numbers were specified as \(K = 2,\ldots,6\). For each candidate, the eigenvectors corresponding to the \(K\) smallest Laplacian eigenvalues were used to construct a spectral embedding. Rows of the embedding were L2-normalized, and cluster assignments were obtained using deterministic discretisation rather than K-means.

The eigengap for each candidate was defined as:

\begin{equation}
\label{eq:supp-9}
g_{K} = \lambda_{K + 1} - \lambda_{K}
\end{equation}
where the eigenvalues were arranged in ascending order. Larger eigengaps indicated stronger spectral separation at the corresponding cluster number.

Permutation-based evaluation of the candidate solutions and subsequent cluster stability analyses are described separately in the Statistical analysis and robustness assessment section.

\subsection{Neurophysiological profiling}\label{supplementary-methods-neurophysiological-profiling}

\subsubsection{Neurophysiological profiling measures}\label{neurophysiological-profiling-measures}

Neurophysiological profiling was performed using cluster assignments obtained from the preceding clustering analysis. These assignments remained fixed, and healthy controls were used exclusively as a reference population.

A total of 106 HbO-derived features were used for primary profiling: 22 regional mean-normalised ALFF (mALFF) values, 66 activation features comprising the condition-specific coefficients for naming, phonemic fluency and semantic fluency (22 ROIs \(\times\) 3 conditions; note that clustering used contrast-based features, as described in the View 2 section above) and 18 anatomical functional connectivity (FC) summaries (3 contexts \(\times\) 3 edge groupings \(\times\) 2 summary statistics).

The FC summaries were derived from three contexts: resting-state connectivity, naming-minus-rest connectivity changes and fluency-minus-rest connectivity changes. For each context, connections were grouped into left intrahemispheric, right intrahemispheric and interhemispheric edges. Two summaries were calculated for each group: the mean signed Fisher-\emph{z} value and the mean absolute value. For rest-to-task contrasts, absolute summaries represented the mean absolute edge-wise FC change, rather than absolute connectivity strength.

In addition, 693 individual FC features (231 per context) were retained for exploratory edge-level analyses. Regional mALFF was interpreted as a relative distribution of low-frequency activity because of its within-participant normalisation.

\subsubsection{Healthy-control-referenced and between-subgroup comparisons}\label{healthy-control-referenced-and-between-subgroup-comparisons}

Each patient cluster was compared with healthy controls using mean differences, Hedges' \emph{g} and signed area under the receiver operating characteristic curve ($\text{AUC}_{\text{signed}}$). The signed AUC was defined as:
\begin{equation}
\label{eq:supp-10}
\mathrm{\text{AUC}_{\text{signed}} = 2 AUC - 1}
\end{equation}
which expresses the direction and magnitude of the cluster--control separation on a scale ranging from -1 to 1.

Positive and negative values indicated that the patient cluster tended to have higher or lower measurements than controls, respectively. Hedges' \emph{g} was calculated as the standardised mean difference with a small-sample correction. 95\% confidence intervals were estimated using 2,000 bootstrap resamples drawn independently within each group.

Cluster-control comparisons used Mann--Whitney U tests. Differences across patient clusters were assessed using Kruskal--Wallis tests, followed by pairwise Mann--Whitney U comparisons.

Benjamini--Hochberg false discovery rate (FDR) correction was applied separately across 318 cluster-control comparisons, 106 patient-cluster omnibus tests and 318 patient-cluster pairwise comparisons. Exploratory edge-level analyses were treated as separate multiple-testing families.

\subsubsection{Cluster-characteristic and shared neurophysiological patterns}\label{cluster-characteristic-and-shared-neurophysiological-patterns}

Representative neurophysiological features were classified according to predefined evidence criteria.

\emph{Cluster-characteristic features}~required FDR-adjusted \emph{q} \textless{} 0.05 for the target cluster versus controls, the overall patient-cluster comparison and both comparisons between the target cluster and the other patient clusters.

\emph{Cluster-differentiating features}~required significant overall and both target pairwise patient-cluster comparisons (\emph{q} \textless{} 0.05), a consistent direction of difference, a minimum absolute pairwise signed AUC of 0.35 and a bootstrap target-extremeness recurrence of at least 0.80. Control-referenced significance was not required. Target-extremeness recurrence quantified the proportion of bootstrap samples in which the target cluster remained the highest or lowest among patient clusters and differed from controls in the same direction.

\emph{Exploratory descriptive patterns}~comprised features not meeting either of the above criteria that were retained as descriptive candidates when bootstrap cluster-control AUC direction consistency was at least 0.95 and either the absolute cluster-control signed AUC was at least 0.50 or patient-cluster separation met the predefined directional effect criterion. These features were not regarded as statistically supported cluster-specific abnormalities.

Shared deviations required consistent directions and FDR-adjusted \emph{q} \textless{} 0.05 for every patient cluster versus controls. Directionally consistent but insufficiently supported changes were classified as trends, while opposing cluster directions were examined as heterogeneous patterns.

\subsubsection{Sensitivity analyses and neurophysiological visualisation}\label{sensitivity-analyses-and-neurophysiological-visualisation}

Bootstrap resampling and leave-one-out analyses assessed the directional consistency and individual influence of representative features, with cluster assignments held fixed.

To assess sensitivity to within-participant mALFF normalisation, a centred log-ratio (CLR) transformation was applied to the 22 positive mALFF values:

\begin{equation}
\label{eq:supp-11}
\mathrm{CLR}_{ij} = \ln\left( \mathrm{mALFF}_{ij} \right) - \operatorname{mean}_{k}\left\lbrack \ln\left( \mathrm{mALFF}_{ik} \right) \right\rbrack
\end{equation}
where \emph{i} indexes participants and \emph{j} and \emph{k} index ROIs. Cluster-control and between-cluster comparisons were repeated using CLR-transformed values. FDR correction for this analysis was performed within the separate mALFF-only testing families.

Demographic sensitivity analyses were conducted within the patient cohort (\emph{n} = 79) using age-adjusted and age-and-sex-adjusted models with rank-based inverse-normal transformed outcomes and HC3 robust covariance estimates.

Primary profiling used all 79 patients with fixed cluster assignments.

\begin{figure}[p]
\centering
\includegraphics[width=\textwidth,height=0.75\textheight,keepaspectratio]{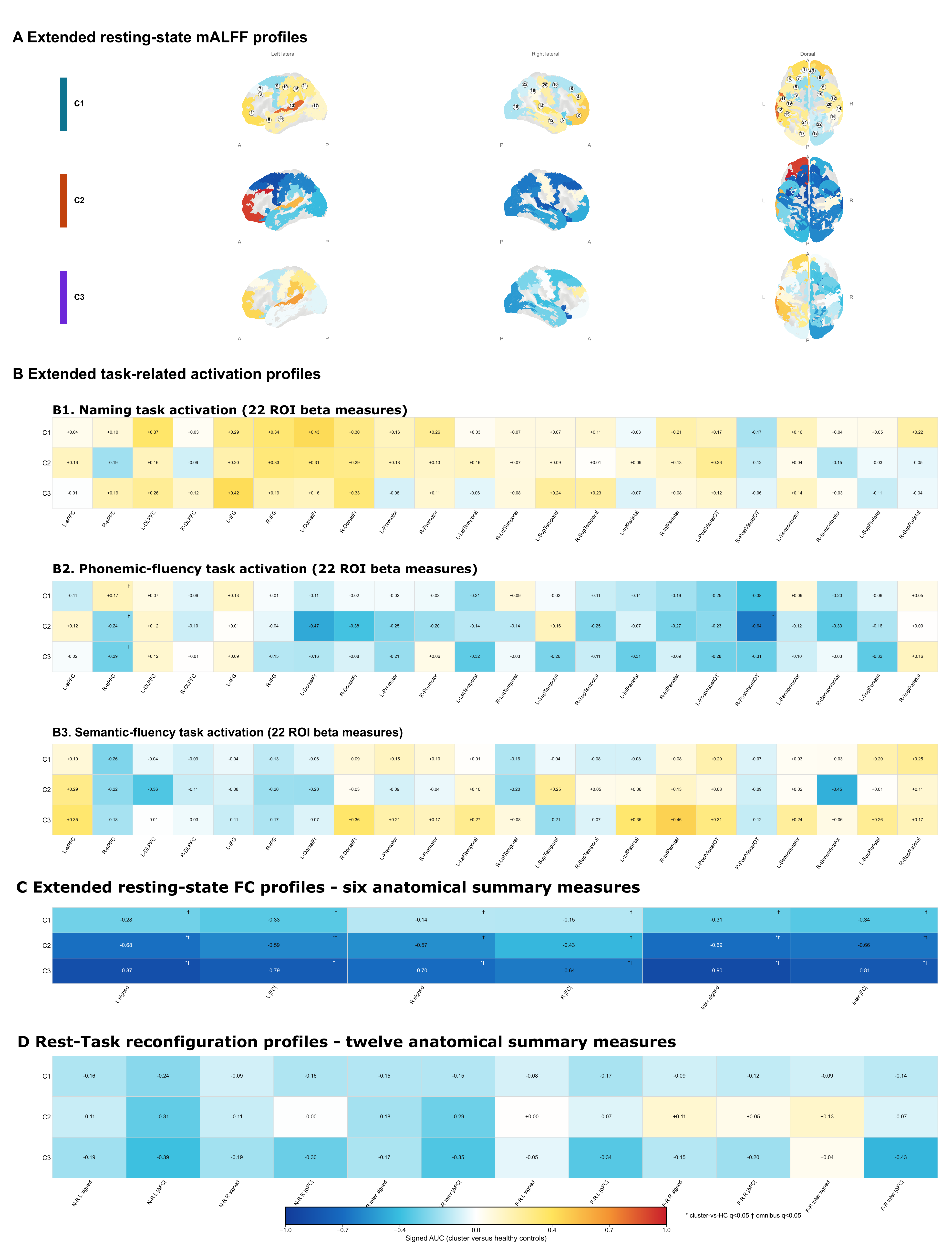}
\caption{\textbf{Extended neurophysiological profiles across the three patient clusters.} The analysis included 79 patients (C1,~\emph{n}~= 32; C2,~\emph{n}~= 17; C3,~\emph{n}~= 30) and 17 healthy controls.~\textbf{(A)}~Resting-state mean amplitude of low-frequency fluctuations (mALFF) across 22 anatomical regions of interest (ROIs), displayed on fixed Talairach Brodmann-area parcels projected through the fsaverage5 cortical ribbon. Parcel geometry is identical across cluster rows. Atlas parcels provide anatomical orientation only and do not represent the spatial extent of fNIRS sensitivity, measurement coverage or vertex-wise statistical effects. Because mALFF was normalised to each participant's whole-ROI mean, colours indicate relative regional alterations.~\textbf{(B)}~The 66 task-activation measures, comprising 22 ROI-level \(\beta\) estimates for naming, phonemic fluency and semantic fluency.~\textbf{(C)}~The six resting-state functional connectivity (FC) summary measures.~\textbf{(D)}~The 12 rest--task reconfiguration summaries, comprising six naming-minus-rest and six fluency-minus-rest measures. Heat-map cells show ROI- or summary-level signed AUC ($\text{AUC}_\text{signed}$) for each cluster relative to healthy controls on a single fixed scale from -1 to +1. An asterisk marks cluster-versus-control~\emph{q}~\textless{} 0.05 and a dagger marks omnibus~\emph{q}~\textless{} 0.05, both adjusted across the global testing family. Non-significant measures are retained. FC and reconfiguration summaries aggregate prespecified edge sets and do not represent edge-specific inference. The figure describes the extended profiles of the clusters and is not an independent validation of clustering.}
\label{fig:supp-profiles}
\end{figure}

Regional profiles were displayed using ROI-based maps and heatmaps, whereas connectivity profiles were visualised using anatomical summaries and FC matrices. All spatial maps were restricted to the sampled fNIRS regions. Extended cluster profiles are presented in Figure~\ref{fig:supp-profiles}.

\subsection{Statistical analysis}\label{supplementary-methods-statistical-analysis}

\subsubsection{Permutation testing for cluster-number evaluation}\label{permutation-testing-for-cluster-number-evaluation}

Statistical support for the candidate number of clusters was evaluated using a feature-permutation procedure with a structured null model. A total of 5,000 surrogate datasets were generated under a structured null model. The 22 mALFF features were permuted jointly across patients, preserving their within-participant multivariate structure, whereas the remaining 759 features were independently permuted column-wise. For each permutation, the complete feature-scaling, similarity network fusion (SNF) and spectral clustering pipeline was repeated. Empirical upper-tail \emph{P} values were calculated from the resulting null distributions of the eigengap for each candidate solution (\emph{K} = 2--6). Holm correction was subsequently applied across the five candidate cluster numbers.

\subsubsection{Statistical analysis of neurophysiological profiles}\label{statistical-analysis-of-neurophysiological-profiles}

Neurophysiological profiling was based on the fixed \emph{K} = 3 patient partition (\emph{N} = 79) and the complete-case healthy control reference cohort (\emph{N} = 17). A total of 106 features were examined.

For each feature, cluster--control differences were evaluated using the Mann--Whitney U test, with signed area under AUC as the effect size.

Differences among the three patient clusters were assessed using Kruskal--Wallis tests, followed by pairwise Mann--Whitney U tests. Benjamini--Hochberg false discovery rate (FDR) correction was performed separately for the following testing families: 318 cluster--control comparisons (106 features \(\times\) 3 clusters), 106 omnibus comparisons among patient clusters and 318 pairwise comparisons among patient clusters (106 features \(\times\) 3 cluster pairs).

\subsubsection{Clustering stability and robustness analyses}\label{clustering-stability-and-robustness-analyses}

The stability of the fixed \emph{K} = 3 partition was evaluated under five complementary perturbation schemes.

\emph{Patient subsampling.}~Eighty per cent of the patients (64 of 79) were randomly sampled without replacement in each iteration. The complete clustering pipeline was repeated 1,000 times, with the number of clusters fixed at three.

\emph{SNF parameter sensitivity.}~The sensitivity of the partition to SNF parameters was assessed using a 5 \(\times\) 4 parameter grid, with the neighbourhood parameter \(K_{\mathrm{SNF}}\)~set to 10, 15, 20, 25 or 30 and the number of iterations \(T\) set to 10, 20, 30 or 50. The primary configuration was \(K_{\mathrm{SNF}}\)~=~20 and \(T\) = 20.

\emph{Leave-one-view-out analysis.}~Each of the three feature views (rest intrinsic organisation, multi-task activation and rest--task reconfiguration) was removed in turn, and the remaining two views were reintegrated using SNF before spectral clustering.

\emph{Alternative fusion baseline.}~The three fixed similarity matrices were combined using direct equal-weight arithmetic averaging as an alternative to SNF. The resulting fused matrix was subjected to the same spectral clustering procedure.

\emph{Leave-one-patient-out analysis.}~Each patient was sequentially removed, and the clustering pipeline was repeated using the remaining 78 patients, resulting in 79 independent refitting runs.

For all perturbation analyses, agreement with the original \emph{K} = 3 partition was quantified using the adjusted Rand index (ARI) and cluster-wise Jaccard indices after Hungarian matching. These analyses assessed internal reproducibility and sensitivity to analytical choices rather than external validation.

\subsubsection{Sensitivity analyses of neurophysiological profiling}\label{sensitivity-analyses-of-neurophysiological-profiling}

Additional sensitivity analyses examined the robustness of the neurophysiological profiles. Bootstrap resampling assessed the recurrence of candidate profile features, and leave-one-patient-out analyses examined the influence of individual participants on the profiling results.

Covariate sensitivity analyses examined the influence of demographic characteristics on between-cluster comparisons, using age-adjusted and age-and-sex-adjusted models with rank-based inverse-normal transformed outcomes and HC3 robust covariance estimates. Because individual-level age and sex information was not available for the healthy-control cohort, covariate adjustment was restricted to comparisons among patient clusters.

Cluster assignments were held fixed throughout these analyses.

\subsubsection{Statistical analysis of clinical and behavioural characteristics}\label{statistical-analysis-of-clinical-and-behavioural-characteristics}

Clinical and behavioural analyses were based on 79 patients with complete baseline records, comprising 32, 17 and 30 patients in the three clusters.

The primary continuous variables were baseline WAB and BDAE scores and the naming, phonemic fluency and semantic fluency task scores. Post-treatment WAB and BDAE scores were excluded from the primary analyses.

Differences among the three patient clusters were assessed using Kruskal--Wallis tests, followed by prespecified two-sided pairwise Mann--Whitney U tests. Holm correction was applied across the five primary omnibus tests and the corresponding 15 pairwise tests.

For task performance, the four groups (healthy controls and Clusters 1--3) were compared using Kruskal--Wallis tests followed by prespecified pairwise Mann--Whitney U tests, with Holm correction applied separately across the three omnibus tests and 18 pairwise comparisons.

Associations between cluster membership and conventional aphasia subtypes were examined using the original, uncollapsed cluster \(\times\) subtype contingency table. Given the sparse expected cell counts, inference was based on a fixed-margin Monte Carlo Pearson chi-square permutation test with 100,000 repetitions. Cramér's \emph{V} was used to quantify the strength of association.

\clearpage
\begingroup
\footnotesize
\setlength{\tabcolsep}{3pt}
\renewcommand{\arraystretch}{1.08}
\begin{longtable}{
  >{\raggedright\arraybackslash}p{0.09\linewidth}
  >{\centering\arraybackslash}p{0.17\linewidth}
  >{\raggedright\arraybackslash}p{0.47\linewidth}
  >{\raggedright\arraybackslash}p{0.16\linewidth}
}
\caption{MNI coordinates and anatomical localisation of the 110 fNIRS channels}
\label{tab:supp-mni-coordinates}\\
\toprule
\textbf{Channel ID} &
\textbf{\shortstack{MNI coordinates\\(x, y, z), mm}} &
\textbf{Anatomical label} &
\textbf{\shortstack{Percentage of\\overlap}} \\
\midrule
\endfirsthead
\multicolumn{4}{c}{\tablename\ \thetable\ (continued)}\\
\toprule
\textbf{Channel ID} &
\textbf{\shortstack{MNI coordinates\\(x, y, z), mm}} &
\textbf{Anatomical label} &
\textbf{\shortstack{Percentage of\\overlap}} \\
\midrule
\endhead
\midrule
\multicolumn{4}{r}{Continued on next page}\\
\endfoot
\bottomrule
\endlastfoot
CH01 & (-29, 65, 16) & BA10 - Frontopolar area & 100\% \\
CH02 & (26, 68, 20) & BA10 - Frontopolar area & 100\% \\
CH03 & (-48, 50, 5) & \begin{minipage}[t]{\linewidth}\raggedright
BA10 - Frontopolar area\\
BA46 - Dorsolateral prefrontal cortex\\
BA47 - Inferior prefrontal gyrus\strut
\end{minipage} & \begin{minipage}[t]{\linewidth}\raggedright
47.97\%\\
39.114\%\\
12.915\%\strut
\end{minipage} \\
CH04 & (-29, 59, 27) & \begin{minipage}[t]{\linewidth}\raggedright
BA9 - Dorsolateral prefrontal cortex\\
BA10 - Frontopolar area\strut
\end{minipage} & \begin{minipage}[t]{\linewidth}\raggedright
16.598\%\\
83.402\%\strut
\end{minipage} \\
CH05 & (26, 61, 29) & \begin{minipage}[t]{\linewidth}\raggedright
BA9 - Dorsolateral prefrontal cortex\\
BA10 - Frontopolar area\strut
\end{minipage} & \begin{minipage}[t]{\linewidth}\raggedright
23.887\%\\
76.113\%\strut
\end{minipage} \\
CH06 & (42, 61, 10) & \begin{minipage}[t]{\linewidth}\raggedright
BA10 - Frontopolar area\\
BA46 - Dorsolateral prefrontal cortex\strut
\end{minipage} & \begin{minipage}[t]{\linewidth}\raggedright
95.968\%\\
4.0323\%\strut
\end{minipage} \\
CH07 & (-53, 35, -9) & BA47 - Inferior prefrontal gyrus & 100\% \\
CH08 & (-48, 45, 19) & \begin{minipage}[t]{\linewidth}\raggedright
BA10 - Frontopolar area\\
BA46 - Dorsolateral prefrontal cortex\strut
\end{minipage} & \begin{minipage}[t]{\linewidth}\raggedright
22.097\%\\
77.903\%\strut
\end{minipage} \\
CH09 & (-28, 52, 37) & \begin{minipage}[t]{\linewidth}\raggedright
BA8 - Includes Frontal eye fields\\
BA9 - Dorsolateral prefrontal cortex\\
BA10 - Frontopolar area\strut
\end{minipage} & \begin{minipage}[t]{\linewidth}\raggedright
1.3453\%\\
84.305\%\\
14.35\%\strut
\end{minipage} \\
CH10 & (26, 56, 36) & \begin{minipage}[t]{\linewidth}\raggedright
BA9 - Dorsolateral prefrontal cortex\\
BA10 - Frontopolar area\strut
\end{minipage} & \begin{minipage}[t]{\linewidth}\raggedright
80.435\%\\
19.565\%\strut
\end{minipage} \\
CH11 & (43, 55, 21) & \begin{minipage}[t]{\linewidth}\raggedright
BA10 - Frontopolar area\\
BA46 - Dorsolateral prefrontal cortex\strut
\end{minipage} & \begin{minipage}[t]{\linewidth}\raggedright
88.235\%\\
11.765\%\strut
\end{minipage} \\
CH12 & (51, 50, -3) & \begin{minipage}[t]{\linewidth}\raggedright
BA10 - Frontopolar area\\
BA11 - Orbitofrontal area\\
BA46 - Dorsolateral prefrontal cortex\\
BA47 - Inferior prefrontal gyrus\strut
\end{minipage} & \begin{minipage}[t]{\linewidth}\raggedright
42.349\%\\
1.4235\%\\
11.032\%\\
45.196\%\strut
\end{minipage} \\
CH13 & (-59, 7, -31) & \begin{minipage}[t]{\linewidth}\raggedright
BA20 - Inferior Temporal gyrus\\
BA21 - Middle Temporal gyrus\\
BA38 - Temporopolar area\strut
\end{minipage} & \begin{minipage}[t]{\linewidth}\raggedright
6.25\%\\
84.191\%\\
9.5588\%\strut
\end{minipage} \\
CH14 & (-57, 22, 2) & \begin{minipage}[t]{\linewidth}\raggedright
BA22 - Superior Temporal Gyrus\\
BA38 - Temporopolar area\\
BA44 - pars opercularis, part of Broca's area\\
BA45 - pars triangularis Broca's area\\
BA47 - Inferior prefrontal gyrus\strut
\end{minipage} & \begin{minipage}[t]{\linewidth}\raggedright
9.4771\%\\
0.65359\%\\
5.5556\%\\
44.118\%\\
40.196\%\strut
\end{minipage} \\
CH15 & (-49, 37, 29) & \begin{minipage}[t]{\linewidth}\raggedright
BA9 - Dorsolateral prefrontal cortex\\
BA46 - Dorsolateral prefrontal cortex\strut
\end{minipage} & \begin{minipage}[t]{\linewidth}\raggedright
11.698\%\\
88.302\%\strut
\end{minipage} \\
CH16 & (-27, 43, 46) & \begin{minipage}[t]{\linewidth}\raggedright
BA8 - Includes Frontal eye fields\\
BA9 - Dorsolateral prefrontal cortex\strut
\end{minipage} & \begin{minipage}[t]{\linewidth}\raggedright
64.574\%\\
35.426\%\strut
\end{minipage} \\
CH17 & (26, 50, 44) & \begin{minipage}[t]{\linewidth}\raggedright
BA8 - Includes Frontal eye fields\\
BA9 - Dorsolateral prefrontal cortex\strut
\end{minipage} & \begin{minipage}[t]{\linewidth}\raggedright
40.889\%\\
59.111\%\strut
\end{minipage} \\
CH18 & (44, 46, 32) & \begin{minipage}[t]{\linewidth}\raggedright
BA9 - Dorsolateral prefrontal cortex\\
BA10 - Frontopolar area\\
BA46 - Dorsolateral prefrontal cortex\strut
\end{minipage} & \begin{minipage}[t]{\linewidth}\raggedright
41.475\%\\
25.346\%\\
33.18\%\strut
\end{minipage} \\
CH19 & (57, 39, 9) & \begin{minipage}[t]{\linewidth}\raggedright
BA10 - Frontopolar area\\
BA45 - pars triangularis Broca's area\\
BA46 - Dorsolateral prefrontal cortex\\
BA47 - Inferior prefrontal gyrus\strut
\end{minipage} & \begin{minipage}[t]{\linewidth}\raggedright
3.01\%\\
28.763\%\\
60.535\%\\
7.6923\%\strut
\end{minipage} \\
CH20 & (52, 28, -14) & \begin{minipage}[t]{\linewidth}\raggedright
BA38 - Temporopolar area\\
BA47 - Inferior prefrontal gyrus\strut
\end{minipage} & \begin{minipage}[t]{\linewidth}\raggedright
29.066\%\\
70.934\%\strut
\end{minipage} \\
CH21 & (-66, -3, -21) & \begin{minipage}[t]{\linewidth}\raggedright
BA20 - Inferior Temporal gyrus\\
BA21 - Middle Temporal gyrus\strut
\end{minipage} & \begin{minipage}[t]{\linewidth}\raggedright
16.495\%\\
83.505\%\strut
\end{minipage} \\
CH22 & (-61, 13, 12) & \begin{minipage}[t]{\linewidth}\raggedright
BA6 - Pre-Motor and Supplementary Motor Cortex\\
BA22 - Superior Temporal Gyrus\\
BA44 - pars opercularis, part of Broca's area\\
BA45 - pars triangularis Broca's area\strut
\end{minipage} & \begin{minipage}[t]{\linewidth}\raggedright
10.458\%\\
3.9216\%\\
61.765\%\\
23.856\%\strut
\end{minipage} \\
CH23 & (-49, 29, 38) & \begin{minipage}[t]{\linewidth}\raggedright
BA8 - Includes Frontal eye fields\\
BA9 - Dorsolateral prefrontal cortex\\
BA46 - Dorsolateral prefrontal cortex\strut
\end{minipage} & \begin{minipage}[t]{\linewidth}\raggedright
13.944\%\\
62.948\%\\
23.108\%\strut
\end{minipage} \\
CH24 & (-25, 39, 51) & \begin{minipage}[t]{\linewidth}\raggedright
BA8 - Includes Frontal eye fields\\
BA9 - Dorsolateral prefrontal cortex\strut
\end{minipage} & \begin{minipage}[t]{\linewidth}\raggedright
97.403\%\\
2.5974\%\strut
\end{minipage} \\
CH25 & (-9, 43, 55) & BA8 - Includes Frontal eye fields & 100\% \\
CH26 & (11, 45, 54) & BA8 - Includes Frontal eye fields & 100\% \\
CH27 & (25, 43, 49) & \begin{minipage}[t]{\linewidth}\raggedright
BA8 - Includes Frontal eye fields\\
BA9 - Dorsolateral prefrontal cortex\strut
\end{minipage} & \begin{minipage}[t]{\linewidth}\raggedright
89.224\%\\
10.776\%\strut
\end{minipage} \\
CH28 & (44, 37, 41) & \begin{minipage}[t]{\linewidth}\raggedright
BA8 - Includes Frontal eye fields\\
BA9 - Dorsolateral prefrontal cortex\\
BA46 - Dorsolateral prefrontal cortex\strut
\end{minipage} & \begin{minipage}[t]{\linewidth}\raggedright
25.571\%\\
65.753\%\\
8.6758\%\strut
\end{minipage} \\
CH29 & (59, 29, 22) & \begin{minipage}[t]{\linewidth}\raggedright
BA9 - Dorsolateral prefrontal cortex\\
BA45 - pars triangularis Broca's area\\
BA46 - Dorsolateral prefrontal cortex\strut
\end{minipage} & \begin{minipage}[t]{\linewidth}\raggedright
1.7123\%\\
40.068\%\\
58.219\%\strut
\end{minipage} \\
CH30 & (59, 14, -4) & \begin{minipage}[t]{\linewidth}\raggedright
BA21 - Middle Temporal gyrus\\
BA22 - Superior Temporal Gyrus\\
BA38 - Temporopolar area\\
BA44 - pars opercularis, part of Broca's area\\
BA45 - pars triangularis Broca's area\\
BA47 - Inferior prefrontal gyrus\strut
\end{minipage} & \begin{minipage}[t]{\linewidth}\raggedright
0.69444\%\\
33.681\%\\
35.069\%\\
1.3889\%\\
4.8611\%\\
24.306\%\strut
\end{minipage} \\
CH31 & (-66, -19, -28) & \begin{minipage}[t]{\linewidth}\raggedright
BA20 - Inferior Temporal gyrus\\
BA21 - Middle Temporal gyrus\strut
\end{minipage} & \begin{minipage}[t]{\linewidth}\raggedright
90.175\%\\
9.8246\%\strut
\end{minipage} \\
CH32 & (-66, -8, 3) & \begin{minipage}[t]{\linewidth}\raggedright
BA21 - Middle Temporal gyrus\\
BA22 - Superior Temporal Gyrus\\
BA42 - Primary and Auditory Association Cortex\strut
\end{minipage} & \begin{minipage}[t]{\linewidth}\raggedright
31.392\%\\
54.693\%\\
13.916\%\strut
\end{minipage} \\
CH33 & (-60, 9, 34) & \begin{minipage}[t]{\linewidth}\raggedright
BA6 - Pre-Motor and Supplementary Motor Cortex\\
BA9 - Dorsolateral prefrontal cortex\\
BA45 - pars triangularis Broca's area\strut
\end{minipage} & \begin{minipage}[t]{\linewidth}\raggedright
36.727\%\\
62.909\%\\
0.36364\%\strut
\end{minipage} \\
CH34 & (-39, 22, 56) & \begin{minipage}[t]{\linewidth}\raggedright
BA6 - Pre-Motor and Supplementary Motor Cortex\\
BA8 - Includes Frontal eye fields\strut
\end{minipage} & \begin{minipage}[t]{\linewidth}\raggedright
7.9167\%\\
92.083\%\strut
\end{minipage} \\
CH35 & (-17, 31, 61) & \begin{minipage}[t]{\linewidth}\raggedright
BA6 - Pre-Motor and Supplementary Motor Cortex\\
BA8 - Includes Frontal eye fields\strut
\end{minipage} & \begin{minipage}[t]{\linewidth}\raggedright
52.471\%\\
47.529\%\strut
\end{minipage} \\
CH36 & (2, 34, 58) & \begin{minipage}[t]{\linewidth}\raggedright
BA6 - Pre-Motor and Supplementary Motor Cortex\\
BA8 - Includes Frontal eye fields\strut
\end{minipage} & \begin{minipage}[t]{\linewidth}\raggedright
31.734\%\\
68.266\%\strut
\end{minipage} \\
CH37 & (18, 35, 59) & \begin{minipage}[t]{\linewidth}\raggedright
BA6 - Pre-Motor and Supplementary Motor Cortex\\
BA8 - Includes Frontal eye fields\strut
\end{minipage} & \begin{minipage}[t]{\linewidth}\raggedright
35.448\%\\
64.552\%\strut
\end{minipage} \\
CH38 & (36, 29, 53) & BA8 - Includes Frontal eye fields & 100\% \\
CH39 & (54, 22, 39) & \begin{minipage}[t]{\linewidth}\raggedright
BA8 - Includes Frontal eye fields\\
BA9 - Dorsolateral prefrontal cortex\\
BA46 - Dorsolateral prefrontal cortex\strut
\end{minipage} & \begin{minipage}[t]{\linewidth}\raggedright
23.022\%\\
71.223\%\\
5.7554\%\strut
\end{minipage} \\
CH40 & (65, 11, 14) & \begin{minipage}[t]{\linewidth}\raggedright
BA6 - Pre-Motor and Supplementary Motor Cortex\\
BA9 - Dorsolateral prefrontal cortex\\
BA22 - Superior Temporal Gyrus\\
BA44 - pars opercularis, part of Broca's area\\
BA45 - pars triangularis Broca's area\strut
\end{minipage} & \begin{minipage}[t]{\linewidth}\raggedright
28.289\%\\
1.3158\%\\
0.65789\%\\
56.908\%\\
12.829\%\strut
\end{minipage} \\
CH41 & (67, -1, -16) & \begin{minipage}[t]{\linewidth}\raggedright
BA20 - Inferior Temporal gyrus\\
BA21 - Middle Temporal gyrus\strut
\end{minipage} & \begin{minipage}[t]{\linewidth}\raggedright
2.027\%\\
97.973\%\strut
\end{minipage} \\
CH42 & (-71, -26, -8) & \begin{minipage}[t]{\linewidth}\raggedright
BA20 - Inferior Temporal gyrus\\
BA21 - Middle Temporal gyrus\\
BA22 - Superior Temporal Gyrus\strut
\end{minipage} & \begin{minipage}[t]{\linewidth}\raggedright
0.31153\%\\
95.639\%\\
4.0498\%\strut
\end{minipage} \\
CH43 & (-68, -13, 25) & \begin{minipage}[t]{\linewidth}\raggedright
BA1 - Primary Somatosensory Cortex\\
BA2 - Primary Somatosensory Cortex\\
BA3 - Primary Somatosensory Cortex\\
BA4 - Primary Motor Cortex\\
BA6 - Pre-Motor and Supplementary Motor Cortex\\
BA40 - Supramarginal gyrus part of Wernicke's area\\
BA43 - Subcentral area\strut
\end{minipage} & \begin{minipage}[t]{\linewidth}\raggedright
15.313\%\\
11.563\%\\
6.875\%\\
11.25\%\\
14.063\%\\
1.25\%\\
39.687\%\strut
\end{minipage} \\
CH44 & (-51, 3, 53) & \begin{minipage}[t]{\linewidth}\raggedright
BA6 - Pre-Motor and Supplementary Motor Cortex\\
BA8 - Includes Frontal eye fields\strut
\end{minipage} & \begin{minipage}[t]{\linewidth}\raggedright
95.618\%\\
4.3825\%\strut
\end{minipage} \\
CH45 & (-29, 15, 65) & \begin{minipage}[t]{\linewidth}\raggedright
BA6 - Pre-Motor and Supplementary Motor Cortex\\
BA8 - Includes Frontal eye fields\strut
\end{minipage} & \begin{minipage}[t]{\linewidth}\raggedright
93.478\%\\
6.5217\%\strut
\end{minipage} \\
CH46 & (-11, 22, 68) & BA6 - Pre-Motor and Supplementary Motor Cortex &
100\% \\
CH47 & (13, 25, 66) & BA6 - Pre-Motor and Supplementary Motor Cortex &
100\% \\
CH48 & (28, 21, 63) & \begin{minipage}[t]{\linewidth}\raggedright
BA6 - Pre-Motor and Supplementary Motor Cortex\\
BA8 - Includes Frontal eye fields\strut
\end{minipage} & \begin{minipage}[t]{\linewidth}\raggedright
75.185\%\\
24.815\%\strut
\end{minipage} \\
CH49 & (48, 16, 54) & \begin{minipage}[t]{\linewidth}\raggedright
BA6 - Pre-Motor and Supplementary Motor Cortex\\
BA8 - Includes Frontal eye fields\strut
\end{minipage} & \begin{minipage}[t]{\linewidth}\raggedright
38.43\%\\
61.57\%\strut
\end{minipage} \\
CH50 & (65, 4, 32) & \begin{minipage}[t]{\linewidth}\raggedright
BA6 - Pre-Motor and Supplementary Motor Cortex\\
BA9 - Dorsolateral prefrontal cortex\\
BA45 - pars triangularis Broca's area\strut
\end{minipage} & \begin{minipage}[t]{\linewidth}\raggedright
65.278\%\\
34.375\%\\
0.34722\%\strut
\end{minipage} \\
CH51 & (69, -7, 4) & \begin{minipage}[t]{\linewidth}\raggedright
BA6 - Pre-Motor and Supplementary Motor Cortex\\
BA21 - Middle Temporal gyrus\\
BA22 - Superior Temporal Gyrus\\
BA42 - Primary and Auditory Association Cortex\\
BA43 - Subcentral area\strut
\end{minipage} & \begin{minipage}[t]{\linewidth}\raggedright
1.2821\%\\
26.923\%\\
56.41\%\\
15.064\%\\
0.32051\%\strut
\end{minipage} \\
CH52 & (-66, -45, -20) & \begin{minipage}[t]{\linewidth}\raggedright
BA20 - Inferior Temporal gyrus\\
BA21 - Middle Temporal gyrus\\
BA37 - Fusiform gyrus\strut
\end{minipage} & \begin{minipage}[t]{\linewidth}\raggedright
54.255\%\\
29.078\%\\
16.667\%\strut
\end{minipage} \\
CH53 & (-70, -38, 10) & \begin{minipage}[t]{\linewidth}\raggedright
BA21 - Middle Temporal gyrus\\
BA22 - Superior Temporal Gyrus\\
BA42 - Primary and Auditory Association Cortex\strut
\end{minipage} & \begin{minipage}[t]{\linewidth}\raggedright
6.7901\%\\
87.037\%\\
6.1728\%\strut
\end{minipage} \\
CH54 & (-63, -21, 44) & \begin{minipage}[t]{\linewidth}\raggedright
BA1 - Primary Somatosensory Cortex\\
BA2 - Primary Somatosensory Cortex\\
BA3 - Primary Somatosensory Cortex\\
BA4 - Primary Motor Cortex\\
BA6 - Pre-Motor and Supplementary Motor Cortex\\
BA40 - Supramarginal gyrus part of Wernicke's area\strut
\end{minipage} & \begin{minipage}[t]{\linewidth}\raggedright
22.105\%\\
23.86\%\\
16.842\%\\
9.4737\%\\
25.263\%\\
2.4561\%\strut
\end{minipage} \\
CH55 & (-40, -6, 65) & \begin{minipage}[t]{\linewidth}\raggedright
BA4 - Primary Motor Cortex\\
BA6 - Pre-Motor and Supplementary Motor Cortex\strut
\end{minipage} & \begin{minipage}[t]{\linewidth}\raggedright
2.6923\%\\
97.308\%\strut
\end{minipage} \\
CH56 & (-19, 6, 72) & BA6 - Pre-Motor and Supplementary Motor Cortex &
100\% \\
CH57 & (3, 9, 70) & BA6 - Pre-Motor and Supplementary Motor Cortex &
100\% \\
CH58 & (22, 9, 71) & BA6 - Pre-Motor and Supplementary Motor Cortex &
100\% \\
CH59 & (42, 4, 62) & \begin{minipage}[t]{\linewidth}\raggedright
BA6 - Pre-Motor and Supplementary Motor Cortex\\
BA8 - Includes Frontal eye fields\strut
\end{minipage} & \begin{minipage}[t]{\linewidth}\raggedright
99.63\%\\
0.37037\%\strut
\end{minipage} \\
CH60 & (61, -8, 47) & \begin{minipage}[t]{\linewidth}\raggedright
BA1 - Primary Somatosensory Cortex\\
BA3 - Primary Somatosensory Cortex\\
BA4 - Primary Motor Cortex\\
BA6 - Pre-Motor and Supplementary Motor Cortex\strut
\end{minipage} & \begin{minipage}[t]{\linewidth}\raggedright
1.4815\%\\
12.963\%\\
7.4074\%\\
78.148\%\strut
\end{minipage} \\
CH61 & (70, -18, 18) & \begin{minipage}[t]{\linewidth}\raggedright
BA1 - Primary Somatosensory Cortex\\
BA2 - Primary Somatosensory Cortex\\
BA3 - Primary Somatosensory Cortex\\
BA40 - Supramarginal gyrus part of Wernicke's area\\
BA42 - Primary and Auditory Association Cortex\\
BA43 - Subcentral area\strut
\end{minipage} & \begin{minipage}[t]{\linewidth}\raggedright
6.1162\%\\
4.893\%\\
1.8349\%\\
17.431\%\\
35.168\%\\
34.557\%\strut
\end{minipage} \\
CH62 & (73, -27, -12) & \begin{minipage}[t]{\linewidth}\raggedright
BA20 - Inferior Temporal gyrus\\
BA21 - Middle Temporal gyrus\strut
\end{minipage} & \begin{minipage}[t]{\linewidth}\raggedright
4.3887\%\\
95.611\%\strut
\end{minipage} \\
CH63 & (-65, -57, -7) & \begin{minipage}[t]{\linewidth}\raggedright
BA20 - Inferior Temporal gyrus\\
BA21 - Middle Temporal gyrus\\
BA37 - Fusiform gyrus\strut
\end{minipage} & \begin{minipage}[t]{\linewidth}\raggedright
0.67568\%\\
40.203\%\\
59.122\%\strut
\end{minipage} \\
CH64 & (-66, -47, 28) & \begin{minipage}[t]{\linewidth}\raggedright
BA22 - Superior Temporal Gyrus\\
BA40 - Supramarginal gyrus part of Wernicke's area\strut
\end{minipage} & \begin{minipage}[t]{\linewidth}\raggedright
10.68\%\\
89.32\%\strut
\end{minipage} \\
CH65 & (-53, -30, 59) & \begin{minipage}[t]{\linewidth}\raggedright
BA1 - Primary Somatosensory Cortex\\
BA2 - Primary Somatosensory Cortex\\
BA3 - Primary Somatosensory Cortex\\
BA40 - Supramarginal gyrus part of Wernicke's area\strut
\end{minipage} & \begin{minipage}[t]{\linewidth}\raggedright
19.922\%\\
36.328\%\\
5.8594\%\\
37.891\%\strut
\end{minipage} \\
CH66 & (-30, -16, 74) & \begin{minipage}[t]{\linewidth}\raggedright
BA4 - Primary Motor Cortex\\
BA6 - Pre-Motor and Supplementary Motor Cortex\strut
\end{minipage} & \begin{minipage}[t]{\linewidth}\raggedright
7.6125\%\\
92.388\%\strut
\end{minipage} \\
CH67 & (-11, -7, 77) & BA6 - Pre-Motor and Supplementary Motor Cortex &
100\% \\
CH68 & (14, -5, 76) & BA6 - Pre-Motor and Supplementary Motor Cortex &
100\% \\
CH69 & (35, -11, 71) & BA6 - Pre-Motor and Supplementary Motor Cortex &
100\% \\
CH70 & (56, -19, 58) & \begin{minipage}[t]{\linewidth}\raggedright
BA1 - Primary Somatosensory Cortex\\
BA2 - Primary Somatosensory Cortex\\
BA3 - Primary Somatosensory Cortex\\
BA4 - Primary Motor Cortex\\
BA6 - Pre-Motor and Supplementary Motor Cortex\\
BA40 - Supramarginal gyrus part of Wernicke's area\strut
\end{minipage} & \begin{minipage}[t]{\linewidth}\raggedright
28.571\%\\
17.063\%\\
46.032\%\\
7.1429\%\\
0.79365\%\\
0.39683\%\strut
\end{minipage} \\
CH71 & (70, -31, 34) & \begin{minipage}[t]{\linewidth}\raggedright
BA1 - Primary Somatosensory Cortex\\
BA2 - Primary Somatosensory Cortex\\
BA40 - Supramarginal gyrus part of Wernicke's area\strut
\end{minipage} & \begin{minipage}[t]{\linewidth}\raggedright
5.6426\%\\
13.48\%\\
80.878\%\strut
\end{minipage} \\
CH72 & (72, -40, 1) & \begin{minipage}[t]{\linewidth}\raggedright
BA21 - Middle Temporal gyrus\\
BA22 - Superior Temporal Gyrus\strut
\end{minipage} & \begin{minipage}[t]{\linewidth}\raggedright
47.988\%\\
52.012\%\strut
\end{minipage} \\
CH73 & (-52, -71, -31) & \begin{minipage}[t]{\linewidth}\raggedright
BA19 - V3\\
BA37 - Fusiform gyrus\strut
\end{minipage} & \begin{minipage}[t]{\linewidth}\raggedright
59.574\%\\
40.426\%\strut
\end{minipage} \\
CH74 & (-60, -69, 4) & \begin{minipage}[t]{\linewidth}\raggedright
BA19 - V3\\
BA21 - Middle Temporal gyrus\\
BA37 - Fusiform gyrus\\
BA39 - Angular gyrus, part of Wernicke's area\strut
\end{minipage} & \begin{minipage}[t]{\linewidth}\raggedright
24.028\%\\
13.074\%\\
43.816\%\\
19.081\%\strut
\end{minipage} \\
CH75 & (-59, -58, 39) & \begin{minipage}[t]{\linewidth}\raggedright
BA39 - Angular gyrus, part of Wernicke's area\\
BA40 - Supramarginal gyrus part of Wernicke's area\strut
\end{minipage} & \begin{minipage}[t]{\linewidth}\raggedright
22.222\%\\
77.778\%\strut
\end{minipage} \\
CH76 & (-40, -40, 68) & \begin{minipage}[t]{\linewidth}\raggedright
BA1 - Primary Somatosensory Cortex\\
BA2 - Primary Somatosensory Cortex\\
BA3 - Primary Somatosensory Cortex\\
BA5 - Somatosensory Association Cortex\\
BA40 - Supramarginal gyrus part of Wernicke's area\strut
\end{minipage} & \begin{minipage}[t]{\linewidth}\raggedright
16.981\%\\
32.83\%\\
7.1698\%\\
29.057\%\\
13.962\%\strut
\end{minipage} \\
CH77 & (-17, -24, 78) & \begin{minipage}[t]{\linewidth}\raggedright
BA3 - Primary Somatosensory Cortex\\
BA4 - Primary Motor Cortex\\
BA6 - Pre-Motor and Supplementary Motor Cortex\strut
\end{minipage} & \begin{minipage}[t]{\linewidth}\raggedright
5.3968\%\\
39.048\%\\
55.556\%\strut
\end{minipage} \\
CH78 & (7, -22, 78) & \begin{minipage}[t]{\linewidth}\raggedright
BA4 - Primary Motor Cortex\\
BA6 - Pre-Motor and Supplementary Motor Cortex\strut
\end{minipage} & \begin{minipage}[t]{\linewidth}\raggedright
1.2821\%\\
98.718\%\strut
\end{minipage} \\
CH79 & (27, -21, 75) & \begin{minipage}[t]{\linewidth}\raggedright
BA3 - Primary Somatosensory Cortex\\
BA4 - Primary Motor Cortex\\
BA6 - Pre-Motor and Supplementary Motor Cortex\strut
\end{minipage} & \begin{minipage}[t]{\linewidth}\raggedright
1.9169\%\\
30.671\%\\
67.412\%\strut
\end{minipage} \\
CH80 & (48, -28, 66) & \begin{minipage}[t]{\linewidth}\raggedright
BA1 - Primary Somatosensory Cortex\\
BA2 - Primary Somatosensory Cortex\\
BA3 - Primary Somatosensory Cortex\\
BA4 - Primary Motor Cortex\\
BA6 - Pre-Motor and Supplementary Motor Cortex\\
BA40 - Supramarginal gyrus part of Wernicke's area\strut
\end{minipage} & \begin{minipage}[t]{\linewidth}\raggedright
22.727\%\\
18.939\%\\
37.879\%\\
14.773\%\\
1.5152\%\\
4.1667\%\strut
\end{minipage} \\
CH81 & (65, -44, 43) & BA40 - Supramarginal gyrus part of
Wernicke's area & 100\% \\
CH82 & (69, -51, 10) & \begin{minipage}[t]{\linewidth}\raggedright
BA21 - Middle Temporal gyrus\\
BA22 - Superior Temporal Gyrus\strut
\end{minipage} & \begin{minipage}[t]{\linewidth}\raggedright
31.475\%\\
68.525\%\strut
\end{minipage} \\
CH83 & (63, -57, -17) & \begin{minipage}[t]{\linewidth}\raggedright
BA19 - V3\\
BA20 - Inferior Temporal gyrus\\
BA37 - Fusiform gyrus\strut
\end{minipage} & \begin{minipage}[t]{\linewidth}\raggedright
0.35971\%\\
33.453\%\\
66.187\%\strut
\end{minipage} \\
CH84 & (-47, -83, -16) & \begin{minipage}[t]{\linewidth}\raggedright
BA18 - Visual Association Cortex (V2)\\
BA19 - V3\strut
\end{minipage} & \begin{minipage}[t]{\linewidth}\raggedright
52\%\\
48\%\strut
\end{minipage} \\
CH85 & (-52, -80, 14) & \begin{minipage}[t]{\linewidth}\raggedright
BA19 - V3\\
BA39 - Angular gyrus, part of Wernicke's area\strut
\end{minipage} & \begin{minipage}[t]{\linewidth}\raggedright
82.772\%\\
17.228\%\strut
\end{minipage} \\
CH86 & (-47, -70, 51) & \begin{minipage}[t]{\linewidth}\raggedright
BA7 - Somatosensory Association Cortex\\
BA19 - V3\\
BA39 - Angular gyrus, part of Wernicke's area\\
BA40 - Supramarginal gyrus part of Wernicke's area\strut
\end{minipage} & \begin{minipage}[t]{\linewidth}\raggedright
42.544\%\\
6.5789\%\\
24.561\%\\
26.316\%\strut
\end{minipage} \\
CH87 & (-26, -49, 74) & \begin{minipage}[t]{\linewidth}\raggedright
BA2 - Primary Somatosensory Cortex\\
BA5 - Somatosensory Association Cortex\\
BA7 - Somatosensory Association Cortex\strut
\end{minipage} & \begin{minipage}[t]{\linewidth}\raggedright
5.4237\%\\
53.22\%\\
41.356\%\strut
\end{minipage} \\
CH88 & (-6, -39, 79) & \begin{minipage}[t]{\linewidth}\raggedright
BA3 - Primary Somatosensory Cortex\\
BA4 - Primary Motor Cortex\\
BA5 - Somatosensory Association Cortex\\
BA6 - Pre-Motor and Supplementary Motor Cortex\strut
\end{minipage} & \begin{minipage}[t]{\linewidth}\raggedright
20.9\%\\
37.621\%\\
18.328\%\\
23.151\%\strut
\end{minipage} \\
CH89 & (17, -38, 80) & \begin{minipage}[t]{\linewidth}\raggedright
BA2 - Primary Somatosensory Cortex\\
BA3 - Primary Somatosensory Cortex\\
BA4 - Primary Motor Cortex\\
BA5 - Somatosensory Association Cortex\\
BA6 - Pre-Motor and Supplementary Motor Cortex\strut
\end{minipage} & \begin{minipage}[t]{\linewidth}\raggedright
3.1847\%\\
37.898\%\\
36.943\%\\
14.968\%\\
7.0064\%\strut
\end{minipage} \\
CH90 & (38, -42, 70) & \begin{minipage}[t]{\linewidth}\raggedright
BA1 - Primary Somatosensory Cortex\\
BA2 - Primary Somatosensory Cortex\\
BA3 - Primary Somatosensory Cortex\\
BA4 - Primary Motor Cortex\\
BA5 - Somatosensory Association Cortex\\
BA7 - Somatosensory Association Cortex\strut
\end{minipage} & \begin{minipage}[t]{\linewidth}\raggedright
11.029\%\\
31.618\%\\
9.5588\%\\
0.36765\%\\
44.118\%\\
3.3088\%\strut
\end{minipage} \\
CH91 & (57, -55, 51) & BA40 - Supramarginal gyrus part of
Wernicke's area & 100\% \\
CH92 & (61, -64, 19) & \begin{minipage}[t]{\linewidth}\raggedright
BA19 - V3\\
BA22 - Superior Temporal Gyrus\\
BA39 - Angular gyrus, part of Wernicke's area\\
BA40 - Supramarginal gyrus part of Wernicke's area\strut
\end{minipage} & \begin{minipage}[t]{\linewidth}\raggedright
10.601\%\\
24.028\%\\
64.311\%\\
1.0601\%\strut
\end{minipage} \\
CH93 & (59, -68, -10) & \begin{minipage}[t]{\linewidth}\raggedright
BA19 - V3\\
BA37 - Fusiform gyrus\strut
\end{minipage} & \begin{minipage}[t]{\linewidth}\raggedright
55.109\%\\
44.891\%\strut
\end{minipage} \\
CH94 & (-42, -92, -11) & \begin{minipage}[t]{\linewidth}\raggedright
BA18 - Visual Association Cortex (V2)\\
BA19 - V3\strut
\end{minipage} & \begin{minipage}[t]{\linewidth}\raggedright
89.879\%\\
10.121\%\strut
\end{minipage} \\
CH95 & (-44, -89, 18) & \begin{minipage}[t]{\linewidth}\raggedright
BA18 - Visual Association Cortex (V2)\\
BA19 - V3\\
BA39 - Angular gyrus, part of Wernicke's area\strut
\end{minipage} & \begin{minipage}[t]{\linewidth}\raggedright
10.442\%\\
81.124\%\\
8.4337\%\strut
\end{minipage} \\
CH96 & (-35, -76, 52) & \begin{minipage}[t]{\linewidth}\raggedright
BA7 - Somatosensory Association Cortex\\
BA19 - V3\strut
\end{minipage} & \begin{minipage}[t]{\linewidth}\raggedright
86.087\%\\
13.913\%\strut
\end{minipage} \\
CH97 & (-17, -56, 74) & \begin{minipage}[t]{\linewidth}\raggedright
BA5 - Somatosensory Association Cortex\\
BA7 - Somatosensory Association Cortex\strut
\end{minipage} & \begin{minipage}[t]{\linewidth}\raggedright
11.486\%\\
88.514\%\strut
\end{minipage} \\
CH98 & (7, -50, 75) & \begin{minipage}[t]{\linewidth}\raggedright
BA4 - Primary Motor Cortex\\
BA5 - Somatosensory Association Cortex\\
BA7 - Somatosensory Association Cortex\strut
\end{minipage} & \begin{minipage}[t]{\linewidth}\raggedright
7.6364\%\\
45.818\%\\
46.545\%\strut
\end{minipage} \\
CH99 & (32, -53, 72) & \begin{minipage}[t]{\linewidth}\raggedright
BA5 - Somatosensory Association Cortex\\
BA7 - Somatosensory Association Cortex\strut
\end{minipage} & \begin{minipage}[t]{\linewidth}\raggedright
36.594\%\\
63.406\%\strut
\end{minipage} \\
CH100 & (49, -61, 55) & \begin{minipage}[t]{\linewidth}\raggedright
BA7 - Somatosensory Association Cortex\\
BA40 - Supramarginal gyrus part of Wernicke's area\strut
\end{minipage} & \begin{minipage}[t]{\linewidth}\raggedright
32.751\%\\
67.249\%\strut
\end{minipage} \\
CH101 & (55, -74, 24) & \begin{minipage}[t]{\linewidth}\raggedright
BA19 - V3\\
BA39 - Angular gyrus, part of Wernicke's area\strut
\end{minipage} & \begin{minipage}[t]{\linewidth}\raggedright
14.981\%\\
85.019\%\strut
\end{minipage} \\
CH102 & (55, -76, -6) & \begin{minipage}[t]{\linewidth}\raggedright
BA18 - Visual Association Cortex (V2)\\
BA19 - V3\\
BA37 - Fusiform gyrus\strut
\end{minipage} & \begin{minipage}[t]{\linewidth}\raggedright
11.871\%\\
74.101\%\\
14.029\%\strut
\end{minipage} \\
CH103 & (-35, -98, -7) & \begin{minipage}[t]{\linewidth}\raggedright
BA17 - Primary Visual Cortex (V1)\\
BA18 - Visual Association Cortex (V2)\\
BA19 - V3\strut
\end{minipage} & \begin{minipage}[t]{\linewidth}\raggedright
0.37736\%\\
96.981\%\\
2.6415\%\strut
\end{minipage} \\
CH104 & (-34, -94, 22) & \begin{minipage}[t]{\linewidth}\raggedright
BA18 - Visual Association Cortex (V2)\\
BA19 - V3\strut
\end{minipage} & \begin{minipage}[t]{\linewidth}\raggedright
0.7326\%\\
99.267\%\strut
\end{minipage} \\
CH105 & (-25, -79, 53) & \begin{minipage}[t]{\linewidth}\raggedright
BA7 - Somatosensory Association Cortex\\
BA19 - V3\strut
\end{minipage} & \begin{minipage}[t]{\linewidth}\raggedright
94.894\%\\
5.1064\%\strut
\end{minipage} \\
CH106 & (-8, -67, 68) & BA7 - Somatosensory Association Cortex &
100\% \\
CH107 & (19, -67, 69) & BA7 - Somatosensory Association Cortex &
100\% \\
CH108 & (38, -71, 54) & \begin{minipage}[t]{\linewidth}\raggedright
BA7 - Somatosensory Association Cortex\\
BA40 - Supramarginal gyrus part of Wernicke's area\strut
\end{minipage} & \begin{minipage}[t]{\linewidth}\raggedright
99.099\%\\
0.9009\%\strut
\end{minipage} \\
CH109 & (48, -80, 30) & \begin{minipage}[t]{\linewidth}\raggedright
BA19 - V3\\
BA39 - Angular gyrus, part of Wernicke's area\strut
\end{minipage} & \begin{minipage}[t]{\linewidth}\raggedright
47.325\%\\
52.675\%\strut
\end{minipage} \\
CH110 & (49, -85, -2) & \begin{minipage}[t]{\linewidth}\raggedright
BA18 - Visual Association Cortex (V2)\\
BA19 - V3\strut
\end{minipage} & \begin{minipage}[t]{\linewidth}\raggedright
29.344\%\\
70.656\%\strut
\end{minipage} \\
\end{longtable}

\noindent\begin{minipage}{\linewidth}
\footnotesize
Coordinates are reported in millimetres in Montreal Neurological
Institute (MNI) space. BA, Brodmann area; fNIRS,
functional near-infrared spectroscopy.

Overlap values were converted from the source proportions to
percentages. For channels with multiple anatomical labels, labels and
percentages correspond line by line; totals may differ slightly from
100\% because of rounding.
\end{minipage}
\endgroup

\clearpage
\begingroup
\small
\setlength{\tabcolsep}{4pt}
\renewcommand{\arraystretch}{1.12}
\begin{longtable}{
  >{\centering\arraybackslash}p{0.08\linewidth}
  >{\raggedright\arraybackslash}p{0.37\linewidth}
  >{\raggedright\arraybackslash}p{0.44\linewidth}
}
\caption{Assignment of the 110 fNIRS channels to 22 anatomical ROIs}
\label{tab:supp-roi-assignment}\\
\toprule
\textbf{ROI No.} & \textbf{ROI name} & \textbf{Channel ID(s)} \\
\midrule
\endfirsthead
\multicolumn{3}{c}{\tablename\ \thetable\ (continued)}\\
\toprule
\textbf{ROI No.} & \textbf{ROI name} & \textbf{Channel ID(s)} \\
\midrule
\endhead
\midrule
\multicolumn{3}{r}{Continued on next page}\\
\endfoot
\bottomrule
\endlastfoot
1 & L\_aPFC (left anterior prefrontal cortex) & CH01, CH03, CH04 \\
2 & R\_aPFC (right anterior prefrontal cortex) & CH02, CH05, CH06,
CH11 \\
3 & L\_DLPFC (left dorsolateral prefrontal cortex) & CH08, CH09, CH15,
CH23, CH33 \\
4 & R\_DLPFC (right dorsolateral prefrontal cortex) & CH10, CH17, CH18,
CH19, CH28, CH29, CH39 \\
5 & L\_IFG (left inferior frontal gyrus) & CH07, CH14, CH22 \\
6 & R\_IFG (right inferior frontal gyrus) & CH12, CH20, CH40 \\
7 & L\_DorsalFrontal (left dorsal frontal cortex) & CH16, CH24, CH25,
CH34 \\
8 & R\_DorsalFrontal (right dorsal frontal cortex) & CH26, CH27, CH36,
CH37, CH38, CH49 \\
9 & L\_Premotor (left premotor cortex) & CH35, CH44, CH45, CH46, CH55,
CH56, CH66, CH67, CH77 \\
10 & R\_Premotor (right premotor cortex) & CH47, CH48, CH50, CH57, CH58,
CH59, CH60, CH68, CH69, CH78, CH79 \\
11 & L\_LateralTemporal (left lateral temporal cortex) & CH13, CH21,
CH31, CH42, CH52 \\
12 & R\_LateralTemporal (right lateral temporal cortex) & CH30, CH41,
CH62 \\
13 & L\_SuperiorTemporal (left superior temporal cortex) & CH32, CH53 \\
14 & R\_SuperiorTemporal (right superior temporal cortex) & CH51, CH72,
CH82 \\
15 & L\_InferiorParietal (left inferior parietal cortex) & CH64, CH75,
CH86 \\
16 & R\_InferiorParietal (right inferior parietal cortex) & CH71, CH81,
CH91, CH92, CH100, CH101, CH109 \\
17 & L\_PosteriorVisualOT (left posterior visual occipitotemporal
cortex) & CH63, CH73, CH74, CH84, CH85, CH94, CH95, CH103, CH104 \\
18 & R\_PosteriorVisualOT (right posterior visual occipitotemporal
cortex) & CH83, CH93, CH102, CH110 \\
19 & L\_Sensorimotor (left sensorimotor cortex) & CH43, CH54, CH65,
CH76, CH88 \\
20 & R\_Sensorimotor (right sensorimotor cortex) & CH61, CH70, CH80,
CH89, CH90 \\
21 & L\_SuperiorParietal (left superior parietal cortex) & CH87, CH96,
CH97, CH105, CH106 \\
22 & R\_SuperiorParietal (right superior parietal cortex) & CH98, CH99,
CH107, CH108 \\
\end{longtable}

\noindent\begin{minipage}{\linewidth}
\footnotesize
Each channel from CH01 to CH110 is assigned exactly once. L, left; R,
right; aPFC, anterior prefrontal cortex; DLPFC, dorsolateral prefrontal
cortex; IFG, inferior frontal gyrus; OT, occipitotemporal.
\end{minipage}
\endgroup

\clearpage

\end{document}